\documentclass[a4paper,10pt]{article}
\usepackage[utf8x]{inputenc}
\usepackage{amsthm,amssymb,amsmath}
\usepackage{graphicx}
\usepackage{epsfig}

\newcommand{\oh}{\Omega h^2}
\newcommand{\gev}{\mathrm{GeV}}
\newcommand{\tev}{\mathrm{TeV}}
\newcommand{\sv}{\langle\sigma v\rangle}
\newcommand{\svt}{\sigma v}
\newcommand{\mh}{M_{H_2}}
\newcommand{\mhone}{M_{H_1}}
\newcommand{\sa}{\sin\alpha}
\newcommand{\mdm}{M_\chi}
\newcommand{\sip}{\sigma_{SI}}

\title{Detection prospects of singlet fermionic dark matter}
\author{Sonja Esch\footnote{sonja.esch@uni-muenster.de}, Michael Klasen\footnote{michael.klasen@uni-muenster.de}  ~and 
Carlos E. Yaguna\footnote{carlos.yaguna@uni-muenster.de} \\ 
\it \small Institut f\"ur Theoretische Physik, Universit\"at M\"unster,\\
\it \small Wilhelm-Klemm-Stra\ss e 9, D-48149 M\"unster, Germany}
\date{}  

\begin{document}
\maketitle
\vspace*{-8cm}
\begin{flushright}
\texttt{MS-TP-13-20}
\end{flushright}
\vspace*{7cm}
\begin{abstract}
A singlet fermion which interacts  only with a new singlet scalar provides a viable and minimal scenario that can explain the dark matter. The singlet fermion is the dark matter particle whereas the new scalar mixes with the Higgs boson providing a link between the dark matter sector and the Standard Model. In this paper, we present an updated  analysis of this model focused on its detection prospects. Both, the parity-conserving case and the most general case are considered. First, the full parameter space of the model is analyzed, and the regions compatible with the dark matter constraint are obtained and characterized. Then, the implications of  current and future direct  detection experiments are taken into account. Specifically, we determine the regions of the multidimensional parameter space that are currently excluded and those that are going to be probed by next generation experiments. 
Finally, indirect detection prospects are discussed and the expected signal at neutrino telescopes  is calculated. 
\end{abstract}

\section{Introduction}
The existence of dark matter provides compelling evidence for physics beyond the Standard Model (SM) but does not tell us what this new physics should be. The experimental data so far, which  relies only on the gravitational effects of the dark matter, allows for  many possible solutions to this long standing puzzle. Hopefully, with the current and new generation of experiments, including the LHC as well as  direct and indirect detection experiments, the dark matter particle will be detected and identified. 

In the meantime, it is important to study in detail the different models that can account for the dark matter. Among them, the so-called minimal models, which try to extend the SM in a minimal way so as to accommodate also the dark matter, look particularly promising. By construction they tend to be simple and predictive, making them easier to analyze and to confront with current and future experimental data. One of these  is the singlet fermionic dark matter model, in which the SM is extended with only  two additional fields, one Majorana fermion ($\chi$) and one real scalar ($\phi$), both singlets under the SM gauge group. The fermion is the dark matter candidate and is assumed to be odd under a $Z_2$ discrete symmetry which is necessary to render it stable. The scalar and the SM fields, in contrast, are taken to be even under the $Z_2$. As a result, the only allowed renormalizable interactions of the dark matter particle are of the form $\bar \chi\chi \phi$ (the parity-conserving case) and $\bar \chi\gamma_5\chi \phi$ (the parity-violating case). The mixing of the singlet scalar $\phi$ with the Higgs provides the link  between the dark matter sector and the SM particles. This singlet fermionic model  can also be seen as an UV completion of the fermionic Higgs portal scenario \cite{Patt:2006fw,Kim:2006af,Kanemura:2010sh,Djouadi:2011aa,Pospelov:2011yp}.

Several studies of the singlet fermionic dark matter model have been published in recent years \cite{Kim:2008pp,Baek:2011aa,LopezHonorez:2012kv,Baek:2012uj,Fairbairn:2013uta}. Earlier works \cite{Kim:2008pp,Baek:2011aa} dealt only with the low mass region, which is excluded by current direct detection bounds. In \cite{LopezHonorez:2012kv} the large mass region was also considered and different ways to avoid the direct detection constraints were identified.  In this paper, we present an   updated analysis of this model focused on its detection prospects. Several improvements with respect to previous works have been incorporated. We make a more general analysis of its parameter space including not only the parity-conserving case but also the general case. 
The most recent experimental data regarding dark matter direct detection \cite{Aprile:2012nq} and Higgs searches at the LHC \cite{Aad:2012tfa,Chatrchyan:2012ufa} are taken into account. We  discuss indirect detection signatures and compute, for the first time, the  expected signal at neutrino telescopes \cite{IceCube:2011ab}  from dark matter annihilations in the Sun. Besides, we make an effort to determine, within the multidimensional parameter space of this model, the regions that are compatible with the dark matter constraint, those that are ruled out by current experiments and those that will be probed by future ones, so as to better assess the detection prospects of singlet fermionic dark matter. 

The rest of the paper is organized as follows. In the next section we introduce our notation, present the model  and discuss its free parameters.   Then in section \ref{sec:par} we show our results concerning the parity-conserving case. After performing a random scan of the parameter space, we illustrate the regions compatible with the dark matter constraint by projecting the viable models into different planes. Direct detection bounds are next imposed and the  prospects for future detection are shown to be promising. Indirect detection signals, on the other hand, are found to be extremely suppressed in this parity-conserving case.   Section \ref{sec:gen} is devoted to the general case, which includes parity-violating interactions. In it,  we made use of a new random scan of the parameter space but the analysis follows similar lines as that of the previous section. Direct detection bounds are in this case not as stringent but indirect detection becomes feasible. The expected neutrino flux, in particular, can be large enough to exclude a handful of models. Finally, we present our conclusions in section \ref{sec:con}. 
\section{The model}
\label{sec:mod}
The model we consider is a minimal extension of the SM with one Majorana fermion, $\chi$, and one real scalar field, $\phi$. $\chi$, the dark matter candidate,  is singlet under the SM gauge group and odd under a $Z_2$ symmetry that guarantees its stability --all other particles including $\phi$ are even under the $Z_2$.  Notice that with $\chi$ alone it is not possible to write renormalizable interaction terms for it. That is  why we also introduce $\phi$, which allows to write Yukawa interaction terms of the form $\bar \chi\chi\phi$. Since $\phi$ is even under the $Z_2$, it will mix with the Higgs boson providing a link between the dark matter sector and the SM.

Excluding the kinetic term, the part of the Lagrangian involving the dark matter particle $\chi$ is given, in the mass eigenstate basis, by   
\begin{equation}
 \mathcal{L}_\chi=-\frac12\left(M_\chi \bar\chi\chi+g_s\phi\bar\chi\chi+ig_p\phi\bar\chi\gamma_5\chi\right),
\end{equation}
where $\mdm$ is the dark matter mass, $g_s$ is the scalar coupling and $g_p$ is the pseudo-scalar one. In addition, the scalar potential is  modified and now reads
\begin{align}
 V(\phi,H)=&-\mu_H^2H^\dagger H + \lambda_H (H^\dagger H)^2-\frac{\mu_\phi^2}{2} \phi^2+\frac{\lambda_\phi}{4} \phi^4+\frac{\lambda_4}{2}\phi^2H^\dagger H \nonumber\\
& + \mu_1^3\phi+\frac{\mu_3}{3}\phi^3+\mu\, \phi (H^\dagger H),
\label{eq:V}
\end{align}
where $H$ is the usual SM Higgs doublet that breaks the electroweak symmetry after acquiring a vacuum expectation value. In the unitary gauge we have that $ H=\frac {1}{\sqrt 2} \begin{pmatrix} 0\\ v+h\end{pmatrix}$ and $\langle H\rangle=\frac {1}{\sqrt 2} \begin{pmatrix} 0\\ v\end{pmatrix}$. In principle,  $\phi$ can also acquire a VEV, but it is possible to choose a basis (by shifting the field) in such a way that $\langle\phi\rangle=0$ --see e.g. \cite{Barger:2007im}. This basis, which had not been used in previous analyses, is quite natural and leads to a simplified description of the phenomenology of the model. In it, the parameter $\mu_1$ is not free but given by
$\mu_1^3=\mu v^2/2$ such that $\partial V/\partial \phi=0$ at the minimum. Throughout this paper we will always use this basis. 

The advantages of choosing a basis where $\langle\phi\rangle=0$ become evident by looking at the Lagrangian. The $\phi-H$ mixing term providing the portal between the dark matter sector and the SM, for instance, originates in this basis \emph{only} from the $\mu$ term\footnote{In a different basis the $\lambda_4$ term would also contribute to the mixing.}. In addition, the $\lambda_\phi$ and $\mu_3$ terms can be straightforwardly interpreted respectively  as quartic and trilinear interaction terms for $\phi$. 

The $\mu$ term in (\ref{eq:V}) induces a mixing between $h$ and $\phi$ which gives rise to two scalar mass eigenstates $H_1$ and $H_2$ defined as
\begin{equation}
 H_1= h\,\cos\alpha+\phi\,\sin\alpha,\quad H_2= \phi\,\cos\alpha  -h\,\sin\alpha,
\end{equation}
where $\alpha$ is the mixing angle. Since $\alpha$ plays a crucial role in all processes coupling the dark matter to the SM particles, we will take it as one of the free parameters of this model.  Notice that  for small mixing $H_1$ becomes a SM-like Higgs, so we will require, in agreement with recent measurements at the LHC \cite{Aad:2012tfa,Chatrchyan:2012ufa}, that $\mhone=125~\gev$. To take into account the fact that the Higgs observed at the LHC is pretty much SM-like \cite{Espinosa:2012ir,CMS:yva,ATLAS:2013sla,Ellis:2013lra}, we will use, following previous works \cite{LopezHonorez:2012kv,Fairbairn:2013uta}, the parameter
\begin{equation}
 r_1\equiv \frac{\sigma_{H_1} \mathrm{Br}_{H_1\to X}}{\sigma^{SM}_{H_1} \mathrm{Br}^{SM}_{H_1\to X}}=\cos^4\alpha \frac{\Gamma_{H_1}^{SM}}{\Gamma_{H_1}}
\label{eq:r1}
\end{equation}
which measures the reduction factor in the number of events originating in the decay of a Higgs boson into the final state $X$. Here, $\sigma_{H_1}$ and Br$_{H_1\to X}$ are the $H_1$ production cross section and branching ratio into $X$, respectively, whereas  $\sigma^{SM}_{H_1}$ and Br$^{SM}_{H_1\to X}$ are the same quantities for a SM Higgs. In our analysis, we always impose $r_1>0.9$. This condition translates, given that the decays $H_1\to \chi\chi$ and $H_1\to H_2H_2$ are kinematically forbidden in the region of interest to us, into a constraint on the mixing angle $\alpha$. A related but weaker bound can be obtained from the non-observation of an additional Higgs boson at the LHC. Defining a parameter $r_2$ analogous  to $r_1$ --see equation (\ref{eq:r1})-- but associated with $H_2$ it is found that $r_2=\sin^4\alpha\,\Gamma^{SM}_{H_2}/\Gamma_{H_2}$.  We require that viable models satisfy $r_2<0.1$, in agreement with current data --see e.g.\cite{Chatrchyan:2013yoa}.

\begin{table}[tb]
\centering
\begin{tabular}{|l|c|}
\hline
Parameter & Range \\ \hline
$M_\chi$ & (50,1000) GeV\\
$\mh$ & (150,1000) GeV \\
$g_s$ & $\pm\pi$($10^{-4}$,1) \\
$\alpha$ & $\pi$ ($10^{-5}$,1)\\
$\lambda_4$ & $\pm4\pi$ ($10^{-4}$,1)\\
$\mu_3$ & $\pm$($10^{-4}$,$10^{4}$) GeV
\\ \hline
\end{tabular}
\caption{\small The free parameters of the fermionic singlet dark matter model and the ranges in which they were allowed to vary in our scan.}   
\label{tab}
\end{table}

The free parameters of this model can be taken to be
\begin{equation}
 \mdm,\mh,g_s,g_p,\alpha, \lambda_4, \mu_3, \lambda_\phi.
\end{equation}
From them one can easily obtain the parameters of the  original scalar potential, equation (\ref{eq:V}). We impose $\lambda_H,\lambda_\phi>0$  and $\lambda_4>-2\sqrt{\lambda_\phi\lambda_H}$ so that the  potential is bounded from below. Since $\lambda_\phi$ does not play any role in the dark matter phenomenology (relic density or dark matter detection), we will simply set it to a fixed value, $\lambda_\phi=3$, throughout this paper.  The real number of free parameters is then six for the parity-conserving case and seven in the general case. We also require  $\chi$ to have a relic density, via a freeze-out mechanism in the early Universe, in agreement with current cosmological measurements \cite{Ade:2013lta}. To ensure that our results are precise, we have  implemented this model into micrOMEGAs \cite{Belanger:2013oya} using LanHEP \cite{Semenov:2010qt}. This allows us to compute the dark matter density with high accuracy even in the presence of resonances or thresholds, both of which,  as we will see, play important roles in this model. We use micrOMEGAs also to calculate dark matter observables such as the spin-independent direct detection cross section, $\svt$ or  the neutrino flux from dark matter annihilation in the Sun. In addition to the relic density constraint, we must ensure that the singlet fermion is consistent with current direct and indirect detection experiments. In \cite{LopezHonorez:2012kv} it had been found that compatibility with direct detection bounds can be achieved in three different cases: i) the pseudo-scalar Higgs portal, in which dark matter interacts via the $\bar\chi\gamma_5\chi\phi$ term; ii) the resonant Higgs portal, where the dark matter annihilation cross section is enhanced close to the $H_1$ or $H_2$ resonance; iii) the indirect Higgs portal, in which the dark matter annihilates into the final state $H_2H_2$. Here  we will confirm  these observations, but we will also find that it is possible to find viable models even below the $H_2$ resonance and that the process $\chi\chi\to H_1H_2$ provides a new  kind of indirect Higgs portal that becomes available at lower masses. Besides,  we will examine in more  detail the regions that are excluded by the XENON100 data \cite{Aprile:2012nq} and those that will be probed by XENON1T \cite{Aprile:2012zx}. Finally, we will also consider   the indirect detection prospects of this model and compute the expected signal at neutrino telescopes. 
\section{The parity-conserving case: $g_p=0$}
\label{sec:par}

In this section we consider the parity-conserving case, $g_p=0$, which is the only one studied in most of the previous works regarding singlet fermionic dark matter. In this case, the dark matter annihilation rate is velocity suppressed with  important consequences for dark matter detection. In the early Universe, this suppression implies that larger couplings are required to satisfy the dark matter constraint, with the result that direct detection experiments set strong bounds on the parameter space of the model. In addition,  the present dark matter annihilation rate $\svt$ is quite suppressed, rendering the indirect detection of dark matter extremely challenging. As we will show, $\svt$ is often more suppressed than naively expected from its velocity dependence. 

 \begin{figure}[t]
\begin{center} 
\includegraphics[scale=0.4]{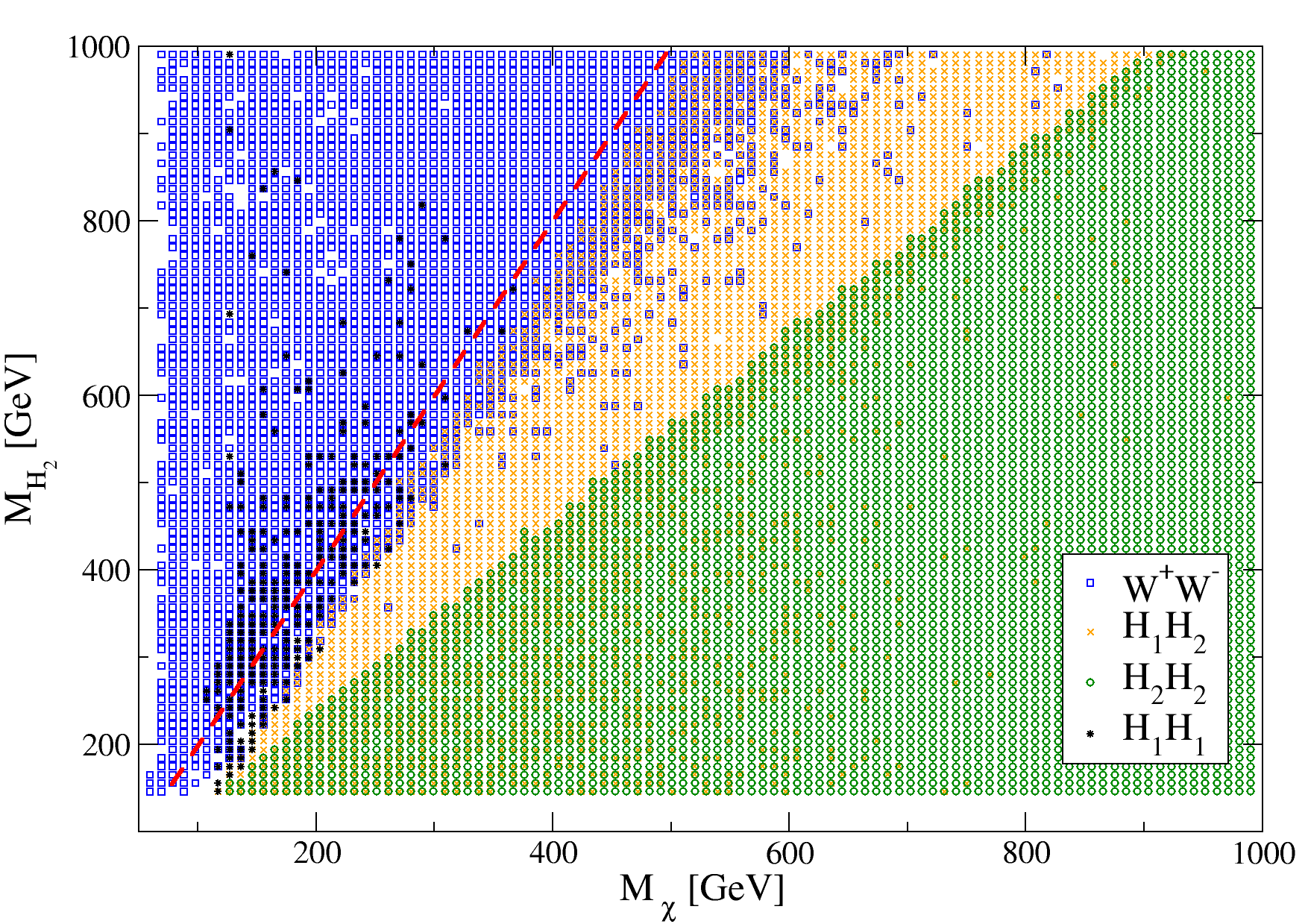}
\caption{\small The region in the plane ($M_\chi$,$\mh$) that is compatible with the dark matter constraint. Different symbols are used to distinguish the dominant annihilation final states. The dashed (red) line shows the resonance condition: $2M_\chi=\mh$.  \label{fig:fig1}}
\end{center}
\end{figure}

Our analysis is based on an extensive random scan of the entire parameter space of this model.  We vary the six free parameters within a very wide range, see  table \ref{tab}. Notice that for definiteness we limit ourselves to particle masses ($\mdm,\mh$) below $1~\tev$. After imposing all the theoretical and experimental bounds mentioned in the previous section, including the relic density constraint, we obtained a sample of about $10^5$ models, on which  the following discussion is based. The resulting viable parameter space is next analyzed and then  the implications of direct and indirect detection experiments are considered.

\subsection{The viable parameter space}

To begin with, let us discuss the viable parameter space. That is the regions that are consistent with the observed dark matter density, $\oh\sim 0.11$. Figure \ref{fig:fig1} projects the viable parameter space  on the plane ($M_\chi$,$\mh$). We see that there are points over the  whole  region we consider, meaning that for any  values of $M_\chi,\mh<1~\tev$, it is always possible to choose the other parameters of the model in such a way that the relic density constraint be satisfied. We also display, in the same figure, the dominant annihilation final states. There are four main possibilities for them: $W^+W^-$ (blue squares), $H_1H_1$ (black stars), $H_1H_2$ (orange exes) and $H_2H_2$ (green circles).  The red dashed line shows the $H_2$-resonant condition, $2M_\chi=\mh$.  In this model, the dark matter annihilation rate can be enhanced at two different resonances, associated with $H_1$ and $H_2$. Since it is the $H_2$-resonance the one that plays a critical role in the phenomenology, we will refer to it simply as the resonance from now on. Below the resonance, the two channels with a final state $H_2$ are closed so dark matter annihilates into SM particles, most models featuring  $W^+W^-$ as the dominant channel but some with $H_1H_1$. Above the resonance new channels open up and quickly dominate the annihilation rate. First we find the $H_1H_2$ final state, whose importance had not been mentioned in previous analyses \cite{LopezHonorez:2012kv,Fairbairn:2013uta}, and  at  larger masses ($\mdm>\mh$) the final state  $H_2H_2$. Hence, in this model the dark matter particle can annihilate into final states with two, one or zero SM particles.

\begin{figure}[t]
\begin{center} 
\includegraphics[scale=0.4]{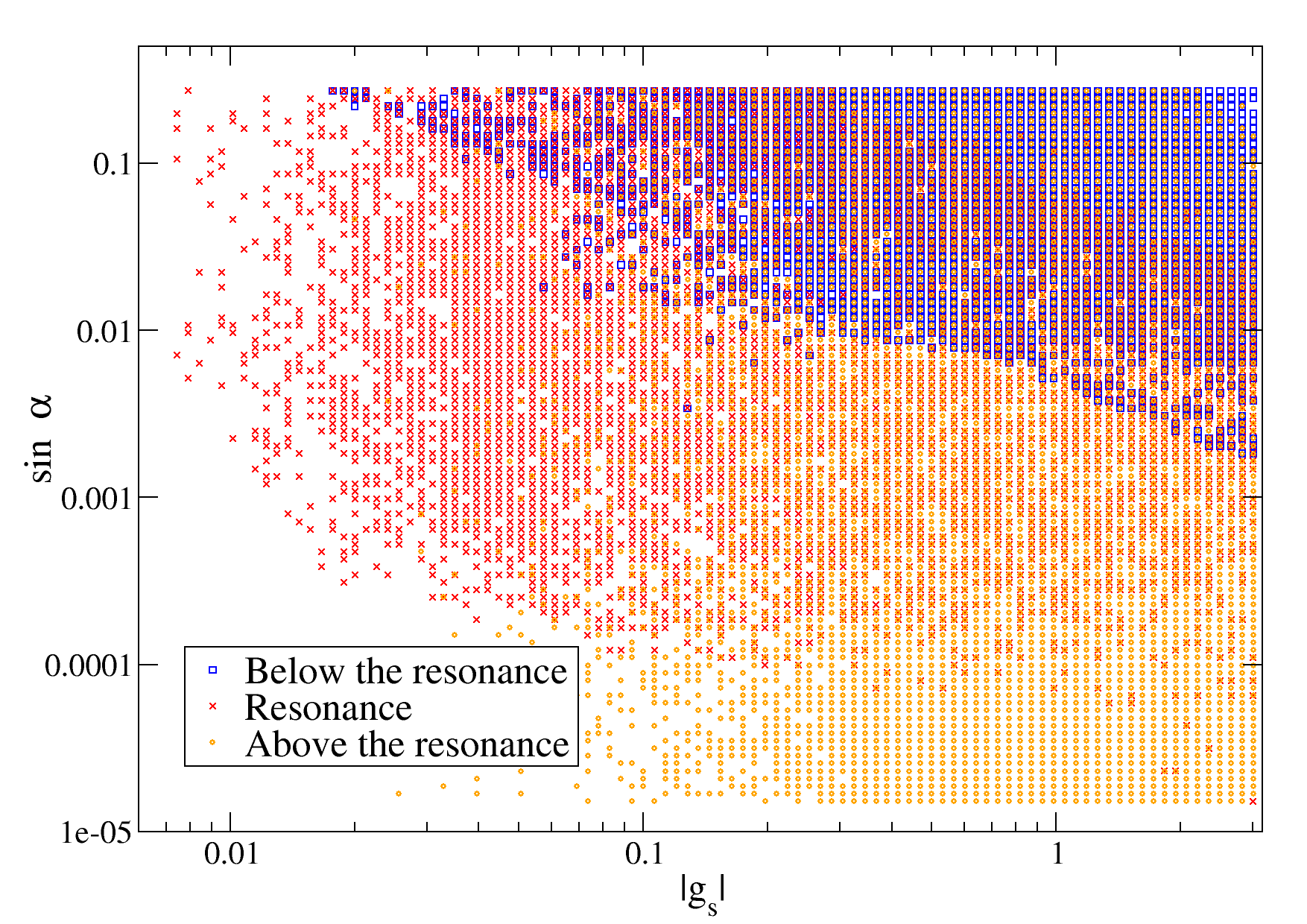}
\caption{\small The region in the plane ($|g_s|$,$\sa$) that is compatible with the dark matter constraint. Different symbols are used to distinguish models below, above and on the resonance. \label{fig:fig3}}
\end{center}
\end{figure}

The annihilation into SM particles, which dominates below the resonance, is controlled by the mixing angle $\alpha$ whereas the annihilations into $H_1H_2$ and $H_2H_2$ receive contributions that do not depend on $\alpha$. We expect, therefore, the dark matter constraint to restrict the possible values of $\alpha$ for models below the resonance.  It is useful to classify the models according to their position with respect to the resonance as \emph{below the resonance} for $2\mdm/\mh<0.9$, \emph{above the resonance} for $2\mdm/\mh>1.1$, and \emph{on the resonance} for  $0.9<2\mdm/\mh<1.1$. In spite of some arbitrariness in the precise definition of the  boundaries for these different regions, we will see that this classification is extremely helpful in understanding qualitatively the phenomenology of the singlet fermionic model. Figure \ref{fig:fig3}, for example, projects  the viable region onto the plane ($|g_s|$,$\sa$) distinguishing the models that are below the resonance (blue squares), above the resonance (orange circles) and on the resonance (red exes). First of all notice that the dark matter coupling is never very small, and that the smallest values are obtained for resonant models.  Most of the non-resonant models feature $g_s>0.1$. This distribution is the result of the dark matter constraint, which forces the annihilation cross section in the early Universe, itself proportional to $g_s^2$,  to be $\sv\sim 3\times 10^{-26}\mathrm{cm^3}/\mathrm{s}$. From the figure we see that for models above the resonance $\sin\alpha$ is unconstrained --it can take any value within the range we examine. Below the resonance, on the other hand, only relatively large values $\sin\alpha\gtrsim 0.002$ are obtained. In fact, all models below the resonance are concentrated in a triangle on the upper right corner of the figure. The reason for this is that, in this case, the cross section goes like the square of  $g_s\sin\alpha$ so this product can never be very small.

\begin{figure}[t]
\begin{center} 
\includegraphics[scale=0.4]{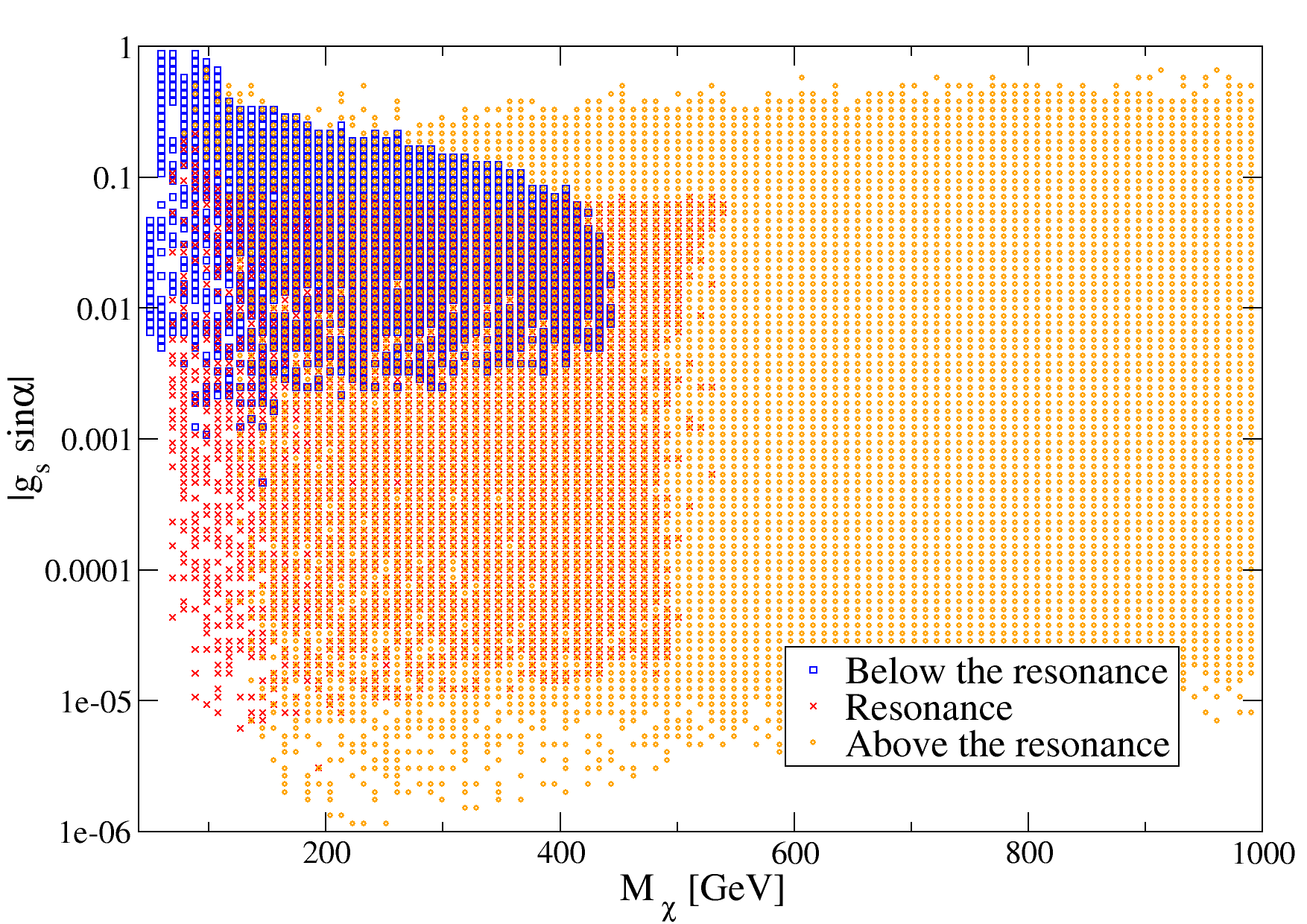}
\caption{\small The region in the plane ($\mdm$, $|g_s\sa|$) that is compatible with the dark matter constraint. Different symbols are used to distinguish models below, above and on the resonance.   \label{fig:fig4}}
\end{center}
\end{figure}

To further elaborate on this point, we display, in figure \ref{fig:fig4}, the viable parameter space in the plane ($M_\chi,|g_s\sin\alpha|$), differentiating models according to their position with respect to the resonance. Due to the range we chose for the parameters, see table \ref{tab}, models below the resonance necessarily  feature $M_\chi<450~\gev$. We learn from the figure that for such models, $|g_s\sin\alpha|\gtrsim 0.001$. Notice also that the lighter the dark matter particle the larger  $|g_s\sin\alpha|$ can be. For models on the resonance, $|g_s\sin\alpha|$  cannot be large, all such models featuring values smaller than $0.1$. Above the resonance, on the other hand, $|g_s\sin\alpha|$ or rather $\sin\alpha$ can take pretty much any value. 

\begin{figure}[t]
\begin{center} 
\begin{tabular}{ccc}
\includegraphics[scale=0.4]{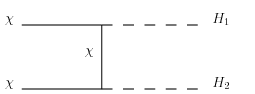} &\includegraphics[scale=0.4]{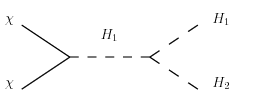} & \includegraphics[scale=0.4]{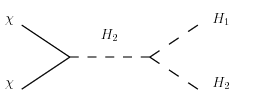}\\[5mm]
(a) & (b) & (c)
\end{tabular}
\caption{\small The Feynman diagrams that contribute to $\chi\chi\to H_1H_2$.   \label{fig:diags}}
\end{center}
\end{figure}

\begin{figure}[t]
\begin{center} 
\includegraphics[scale=0.4]{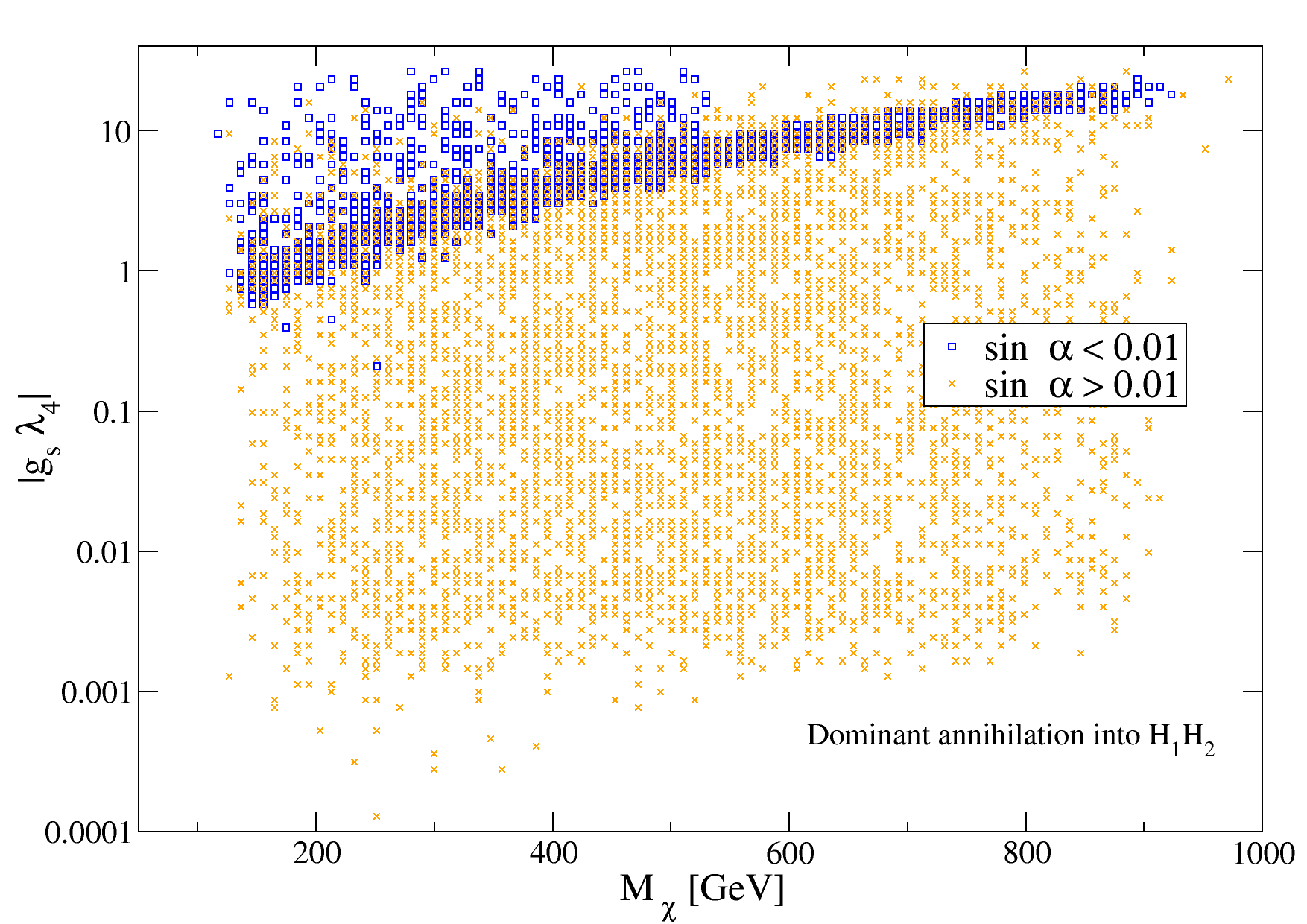}
\caption{\small The region in the plane ($\mdm$, $|g_s\lambda_4|$) for models  compatible with the dark matter constraint in which the dark matter annihilates dominantly into the final state $H_1H_2$. Different symbols are used to distinguish models with values of $\sa$ larger or smaller than $0.01$. \label{fig:fig2}}
\end{center}
\end{figure}

In figure \ref{fig:fig1} we learned of a new viable region  above the resonance where the dominant annihilation final state is $H_1H_2$. Since this possibility seems to have been overlooked in previous studies of this model (e.g. in \cite{LopezHonorez:2012kv}),  we want to describe it in some detail. Three different kind of diagrams contribute to $\chi\chi\to H_1H_2$ as shown in  figure \ref{fig:diags}. The first one, (a), is proportional to $\sin\alpha$, the second, (b), goes like $\sin^2\alpha$, but the last one, (c), is instead determined by the product $g_s\lambda_4$ and does not depend on $\sin\alpha$, allowing, as we will see, to avoid the direct detection bound. The parameter $\lambda_4$, in fact, sets the $H_2H_2H_1$ vertex as can be read from equation (\ref{eq:V}).  Figure \ref{fig:fig2} shows, for models that annihilate dominantly into $H_1H_2$, a scatter plot of   $|g_s\lambda_4|$ versus the dark matter mass. We illustrate the dependence with $\sin\alpha$ by considering two different ranges for it, larger or smaller than $10^{-2}$. Notice that when  $\sin\alpha<0.01$,  $|g_s\lambda_4|$ tends to be large and varies over a narrow range, in agreement with the above discussion. If, on the other hand, $\sin\alpha>0.01$ the contribution from diagrams (a) and (b) becomes relevant and $|g_s\lambda_4|$ can reach much smaller values.

We have in this way determined the regions in the parameter space that are consistent with the dark matter constraint. Next we would like to know how these regions are modified once we impose current dark matter bounds, particularly the direct detection constraints from XENON100. And we would like to determine the capability of future experiments such as XENON1T to probe even further the remaining viable regions.

\subsection{Detection bounds and prospects}

Because direct detection bounds are known to restrict in a significant way the parameter space of this model, it is important to include the most recent experimental bound, that published by the XENON100 collaboration in mid 2012 \cite{Aprile:2012nq}. A novelty of our analysis is that we also examine the prospects for  detection in XENON1T \cite{Aprile:2012zx}, which is expected to start taking data in 2015.

\subsubsection{Direct detection}
\begin{figure}[t]
\begin{center} 
\includegraphics[scale=0.4]{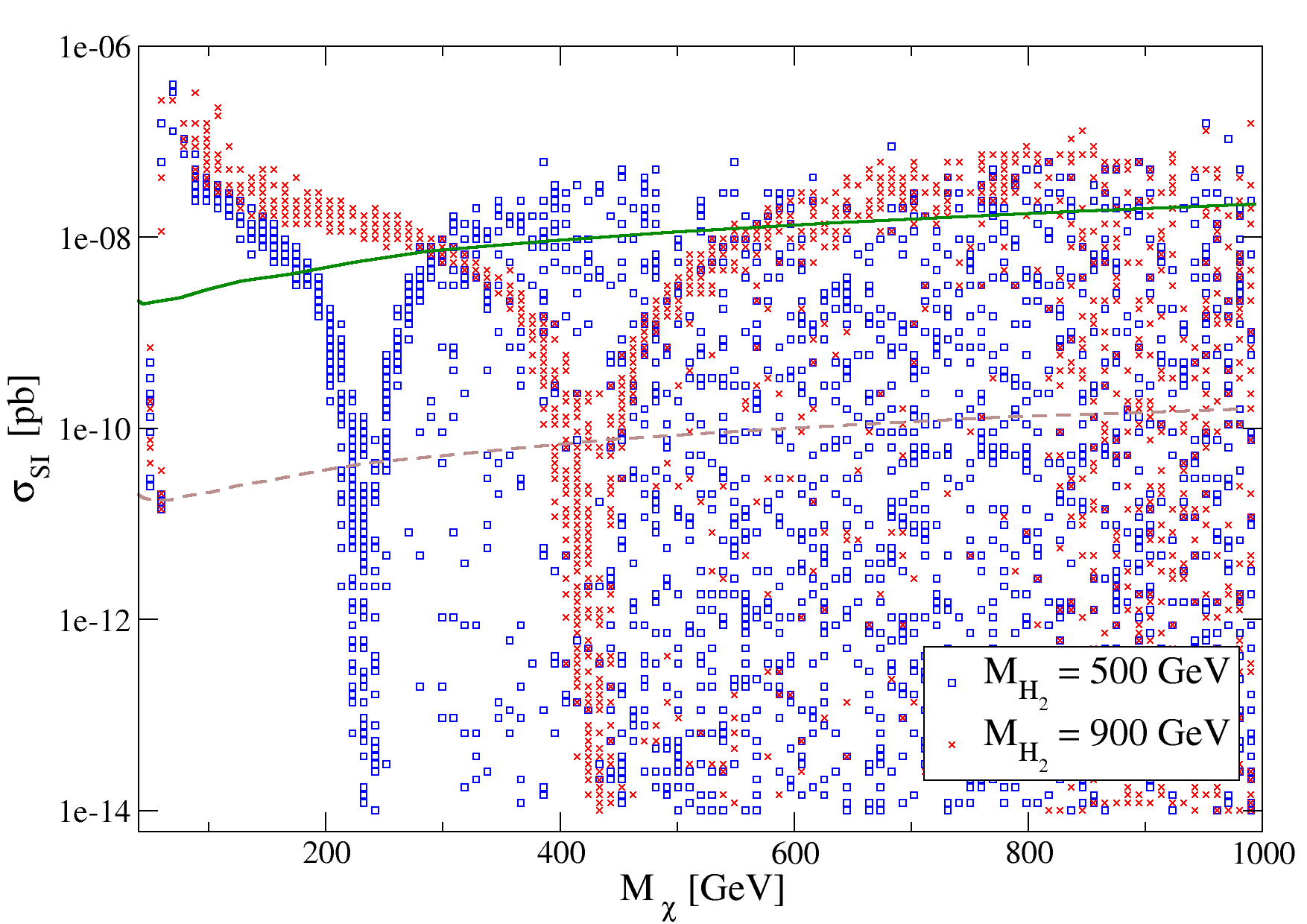}
\caption{\small The spin-independent direct detection cross section as a function of the dark matter mass for two different values of $\mh$: $500~\gev$ (blue squares), $900~\gev$ (red exes). The green solid line shows the current bound from the XENON100 experiment while the dashed line displays the expected sensitivity of XENON1T. \label{fig:fig13}}
\end{center}
\end{figure}

In this model, the scattering of dark matter particles on nuclei is spin-independent and proceeds via $t$-channel exchange of $H_1$ and $H_2$. The elastic scattering cross section $\sip$ of $\chi$ off a proton $p$ is given by 
\begin{equation}
 \sip=\frac{g_s^2\sin^22\alpha}{4\pi}m_r^2\left(\frac{1}{\mhone^2}-\frac{1}{\mh^2}\right)^2 g_{Hp}^2,
\label{eq:dd}
\end{equation}
where $m_r$ is the reduced mass and 
\begin{equation}
 g_{Hp}=\frac{m_p}{v}\left[\sum_{q=u,d,s}f_q^p+\frac 29\left(1-\sum_{q=u,d,s}f_q^p\right)\right]\approx 10^{-3}.
\end{equation}
For the form factors, $f_q^p$, we use the default values from micrOMEGAs. Notice that $\sip$ is essentially independent of the dark matter mass (since $m_r\approx m_p$) and proportional to $g^2\sin^2\alpha$ ($\sin2\alpha\sim 2\sin\alpha$).

\begin{figure}[tb]
\begin{center} 
\includegraphics[scale=0.4]{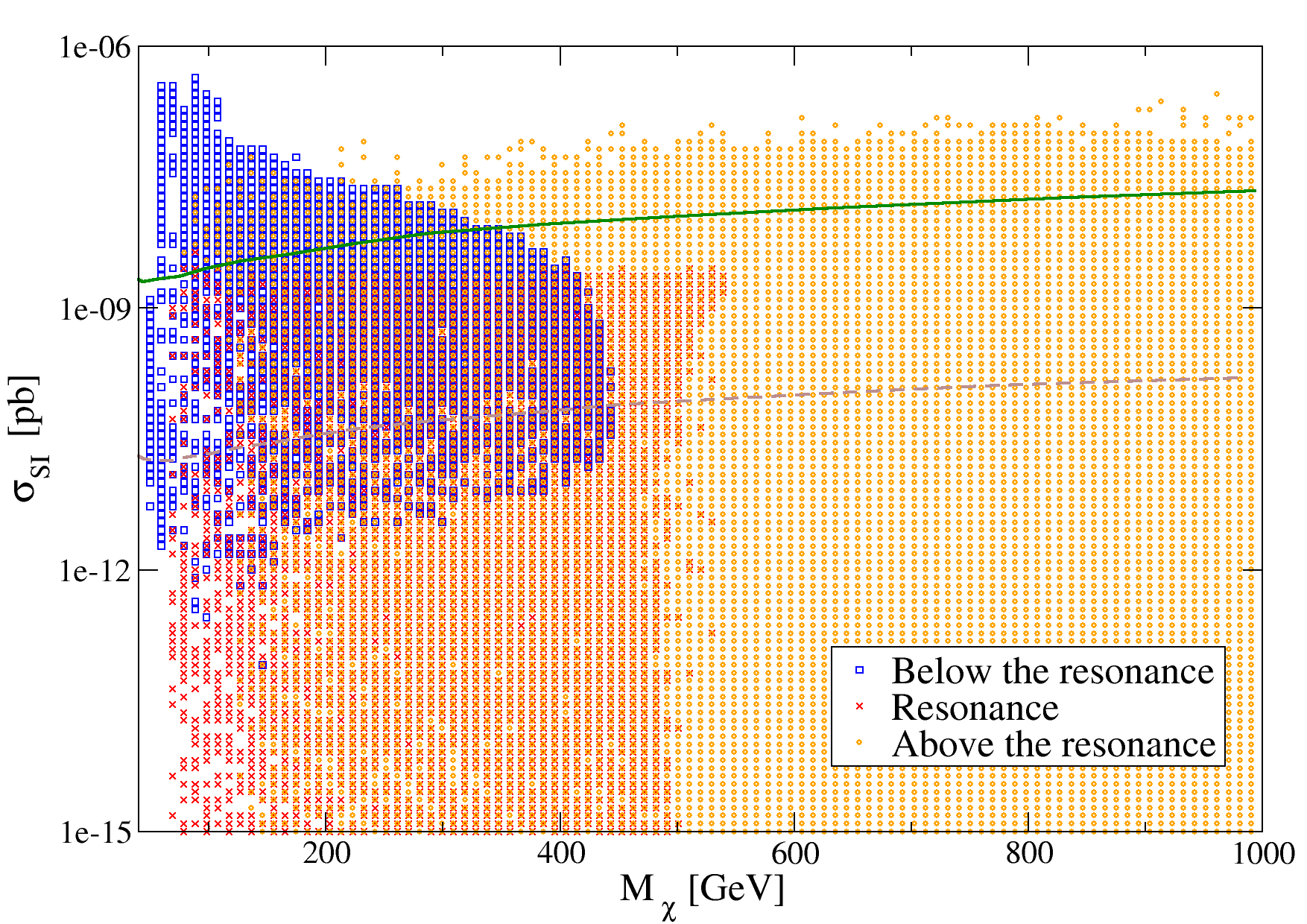}
\caption{\small The spin-independent direct detection cross section as a function of the dark matter mass for our set of models. Different symbols are used to distinguish models below, above and on the resonance. The green solid line shows the current bound from the XENON100 experiment while the dashed line displays the expected sensitivity of XENON1T. \label{fig:fig5}}
\end{center}
\end{figure}

In previous works \cite{LopezHonorez:2012kv,Fairbairn:2013uta}, it had already been shown that the direct detection constraints can be strong enough to exclude significant regions of the parameter space compatible with the dark matter bound. To make contact with such studies, we first illustrate in figure \ref{fig:fig13} the direct detection cross section versus the dark matter mass for fixed  values of $\mh$: $500~\gev$ (blue) and  $900~\gev$ (red). Notice that the low mass region of this figure agrees with the results shown in \cite{Fairbairn:2013uta} (figure 2) but not with those in \cite{LopezHonorez:2012kv} (figure 2). The suppression of the cross section  close to the resonance is clearly observed in the figure. Below the resonance, there is  a correlation between the dark matter density and the scattering cross section so $\sip$ lies in a narrow band.   Above the resonance that correlation is partially lost due to the appearance of the new annihilation final states $H_1H_2$ and $H_2H_2$ with the result that  $\sip$ can  vary over several orders of magnitude.  For comparison we also show as a solid green line the current bound from XENON100 \cite{Aprile:2012nq} and as a dashed-line the expected sensitivity of XENON1T \cite{Aprile:2012zx}. Viable models start at $200~\gev$ for $\mh=500~\gev$ and at about $300~\gev$ for $\mh=900~\gev$.  This figure illustrates  that, even below the resonance, it is possible  to satisfy the dark matter constraint and current direct detection bounds.

\begin{figure}[tb]
\begin{center} 
\begin{tabular}{cc}
\includegraphics[scale=0.2]{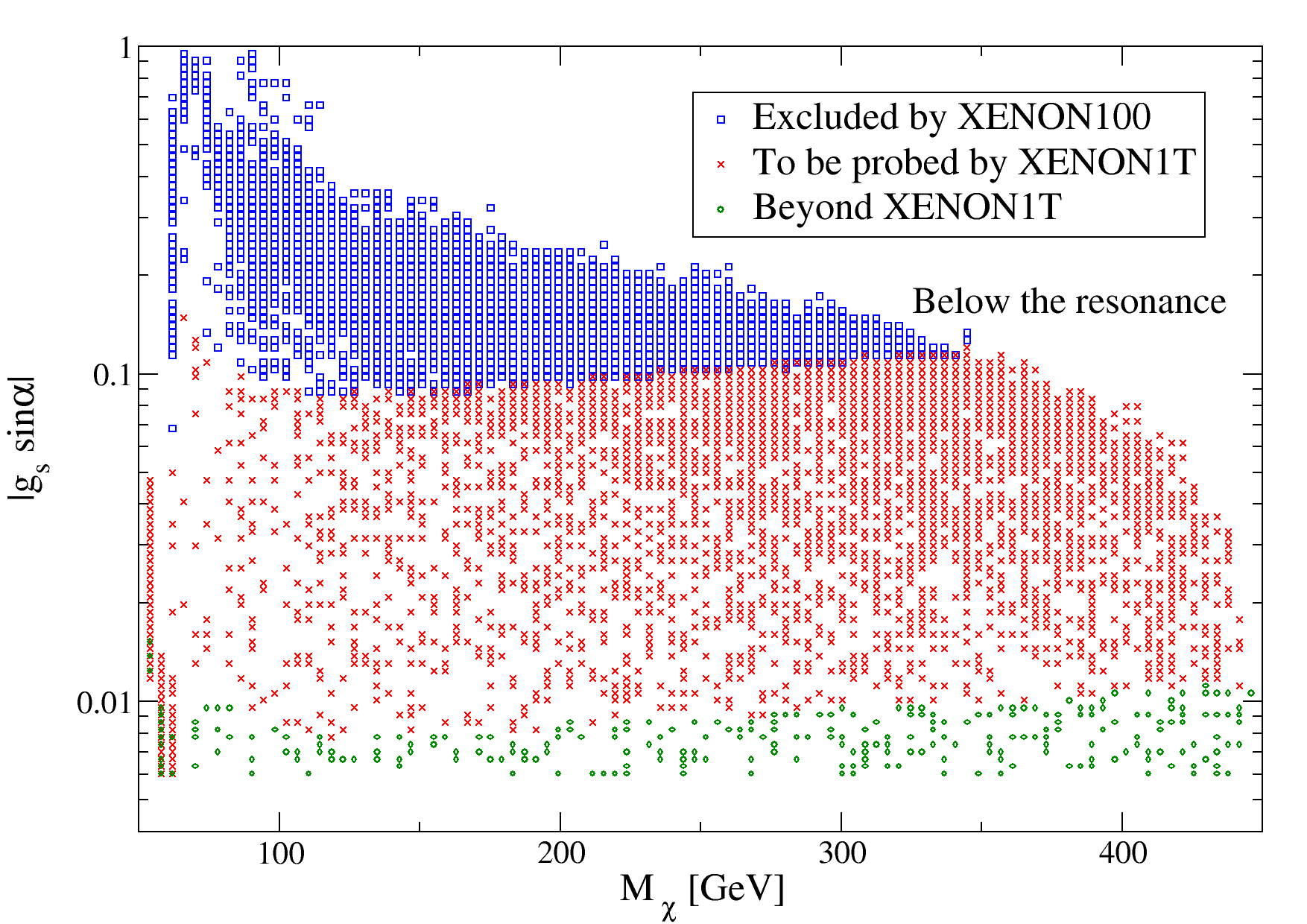} & \includegraphics[scale=0.2]{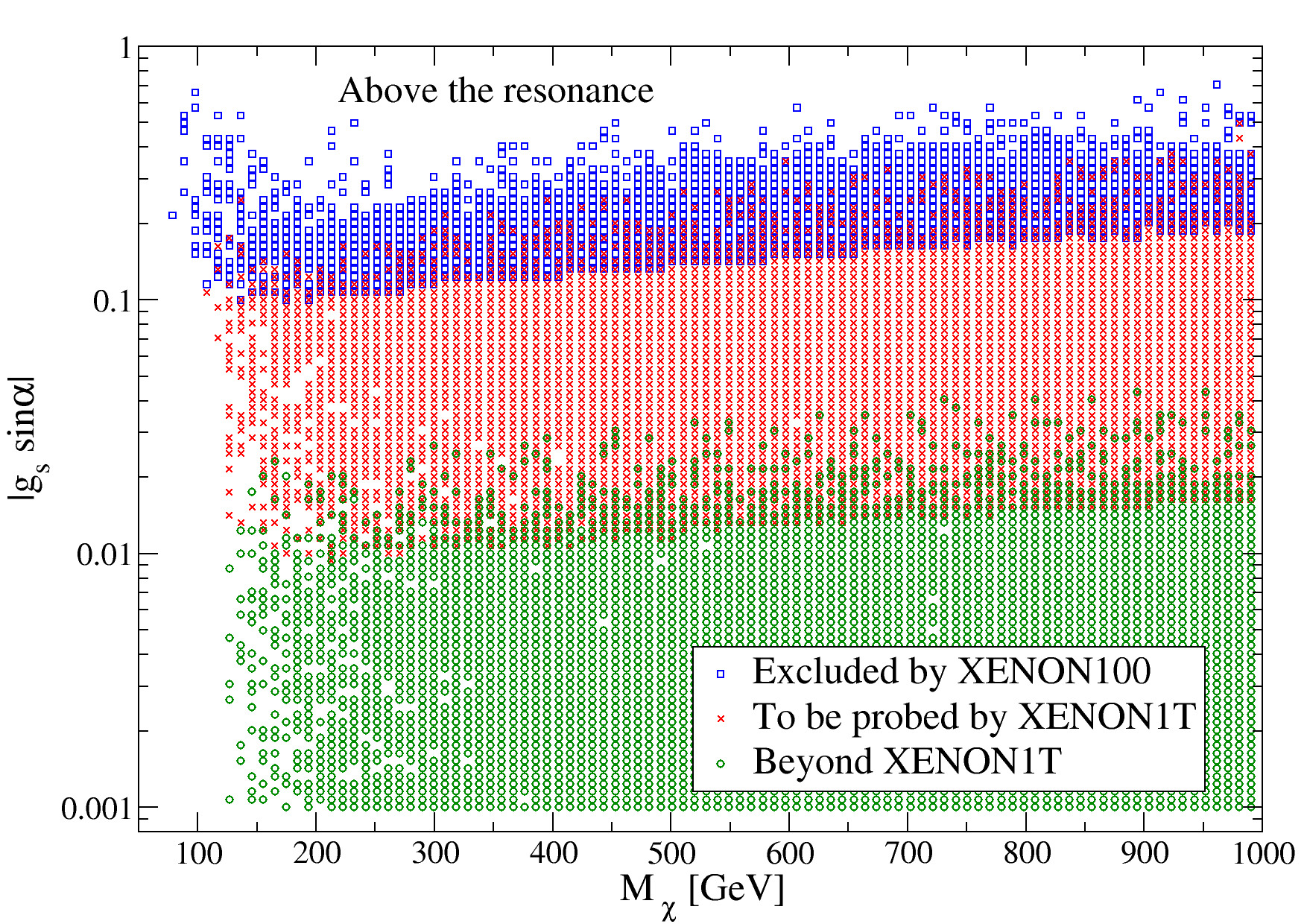}
\end{tabular}
\caption{\small The reach of current and future direct detection experiments  in the plane ($\mdm$,$g_s\sa$) for models below (left) and above (right) the resonance. The convention is as follows: blue squares for models that are currently excluded by the XENON100 bound, red exes for  models with a $\sip$ within the expected sensitivity of XENON1T, and green circles for models with a $\sip$ below the expected sensitivity of XENON1T. \label{fig:fig6/7}}
\end{center}
\end{figure}

\begin{figure}[bt]
\begin{center} 
\includegraphics[scale=0.4]{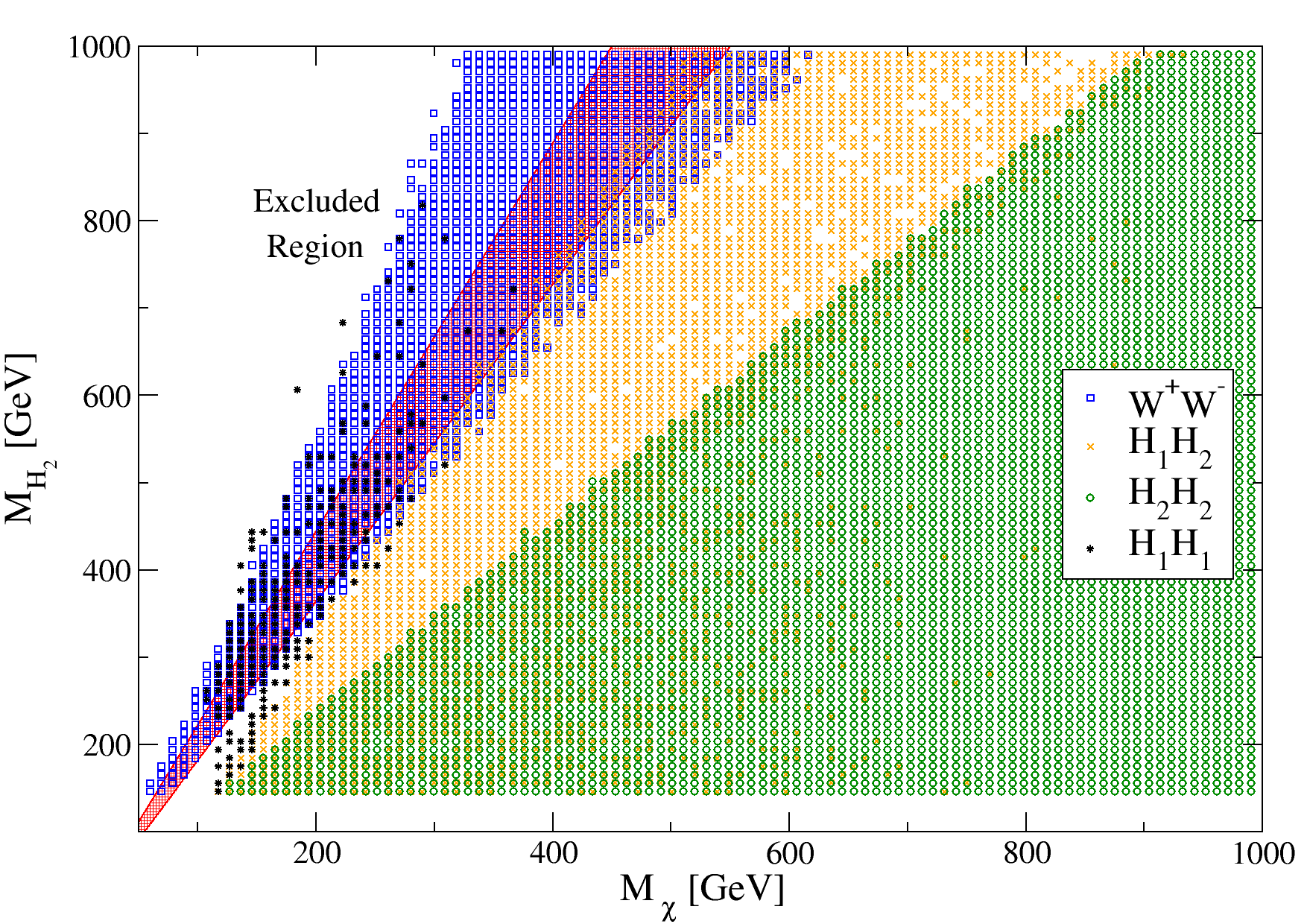}
\caption{\small The region in the plane ($\mdm$,$\mh$) that is compatible with the XENON100 bound and the dark matter constraint.  Different symbols are used to distinguish the dominant annihilation final states. The red band illustrates the resonance region.  \label{fig:fig8}}
\end{center}
\end{figure}

In figure \ref{fig:fig5} we generalize the above results to arbitrary values of $\mh$. Again we classify the models according to their position with respect to the resonance. It is clear that for dark matter masses above the resonance $\sip$ can be very suppressed whereas it is never that small for models below the resonance. As before, we also display the current bound from XENON100 and the expected sensitivity of XENON1T. Notice that one can find models satisfying the current bound for any mass, both below and above the resonance. From the figure we see that pretty much all resonant models (red exes) are able to evade the current bound. Future experiments such as XENON1T will play a crucial role in testing this model. They will for the first time be sensitive to  the resonant models, they will be able to exclude a large fraction of models above the resonance,  and they will probe most of the parameter space of models below the resonance.

Since the direct detection cross section is determined by $g_s\sin\alpha$ --see equation (\ref{eq:dd})-- it is useful to look at the excluded/non-excluded regions in the plane ($\mdm$,$|g_s\sin\alpha|$), as shown in figure \ref{fig:fig6/7}. The left panel, which displays models below the resonance, shows that XENON100 excludes models featuring $|g_s\sin\alpha|\gtrsim 0.1$ (blue squares) and that XENON1T will probe models up to $|g_s\sin\alpha|\sim 0.01$ (red exes). Very few models will be below the expected XENON1T sensitivity (green circles).The right panel shows the same figure for models above the resonance. In this case there is a significant overlap between the different regions. The excluded models lie between $|g_s\sin\alpha|\gtrsim 0.1$ for $\mdm\sim 100~\gev$ and  $|g_s\sin\alpha|\gtrsim 0.2$ for $\mdm\sim 1~\tev$ (blue points).  XENON1T will probe $|g_s\sin\alpha|$  one order of magnitude further (red points) but many models featuring even smaller detection cross sections will  be beyond its expected sensitivity.

To summarize our findings regarding current direct detection bounds, we show in figure \ref{fig:fig8} the regions in the plane ($\mdm$,$\mh$) that are compatible with the dark matter constraint and with the XENON100 bound, classified according to the dominant annihilation channel. In the figure we also show (in red) the region that corresponds to models on the resonance. Models to its left  are below the resonance and those to its right are above the resonance. Comparing it against figure \ref{fig:fig1}, where the XENON100 bound was not imposed, we no longer see  the left wedge of low mass models (which we indicate in the figure by Excluded Region). The bound requires, for example, $\mh\lesssim 500~\gev$ for $\mdm=200~\gev$ and $\mh\lesssim 900~\gev$ for $\mdm=300~\gev$. If $\mdm\gtrsim 350~\gev$, the bound can be satisfied for any value of $\mh$ within the range considered. The bound excludes also many models with large masses and $W^+W^-$ as their dominant annihilation channel. Some models featuring $H_1H_2$ and $H_2H_2$ as their dominant annihilation final states are seen to be excluded but in general they are less affected by this bound. Undoubtedly, the XENON100 bound is quite relevant for singlet fermionic dark matter but it is possible to satisfy it even for models below the resonance.

\subsubsection{Indirect Detection}
\begin{figure}[t]
\begin{center} 
\includegraphics[scale=0.4]{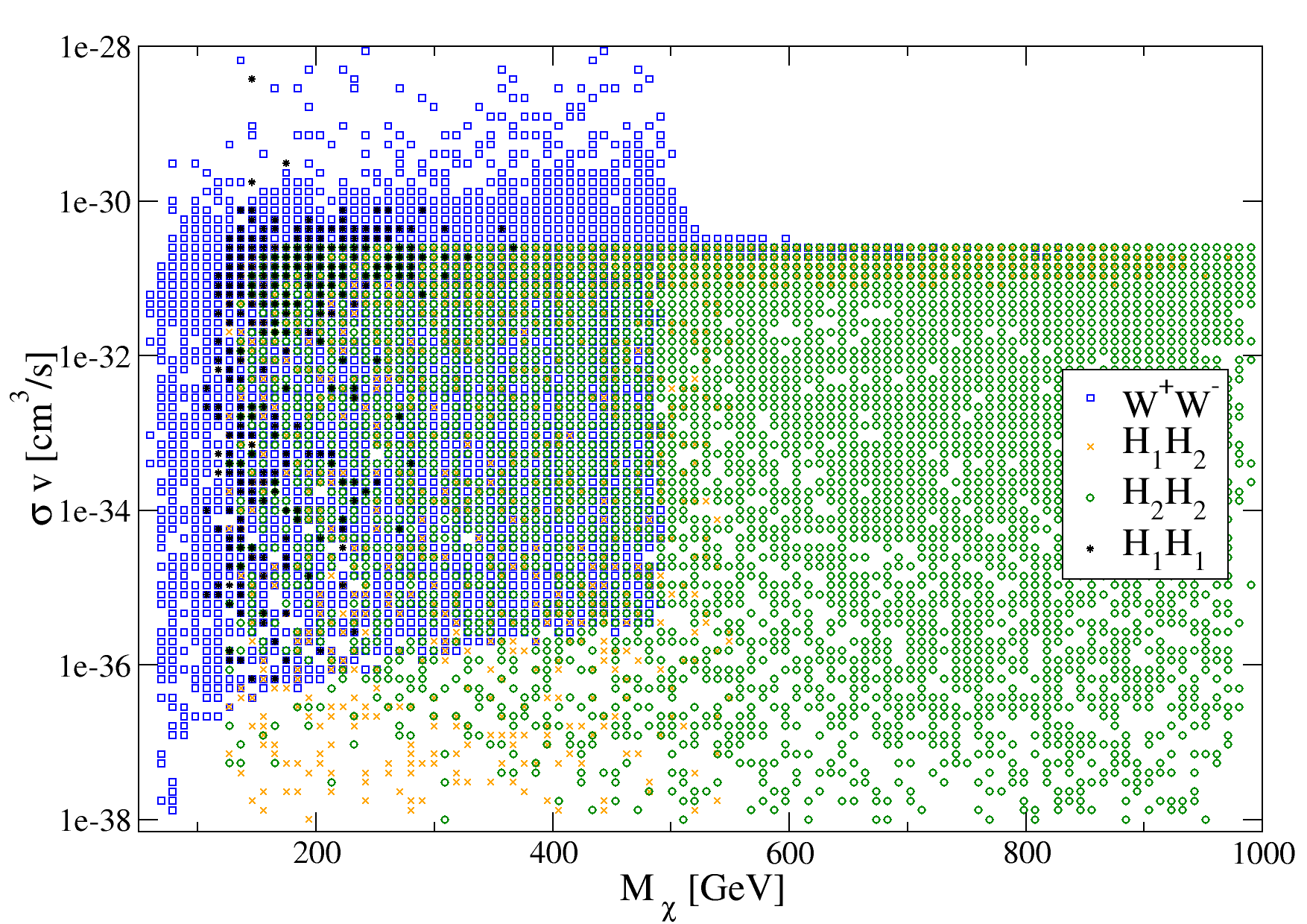}
\caption{\small The dark matter annihilation rate today, $\svt$, versus the dark matter mass for our set of models.  Different symbols are used to distinguish the dominant annihilation final states. \label{fig:fig9}}
\end{center}
\end{figure}

In the parity-conserving case, the dark matter annihilation rate is suppressed by the square of the dark matter velocity, $v$. In the early Universe this suppression is not that  important because around the freeze-out temperature, $v^2\sim 1/20$. But today, in the galactic halo we have $v^2\sim 10^{-6}$, so we generically expect poor prospects for the indirect detection of dark matter in this model.

\begin{figure}[bt]
\begin{center} 
\includegraphics[scale=0.4]{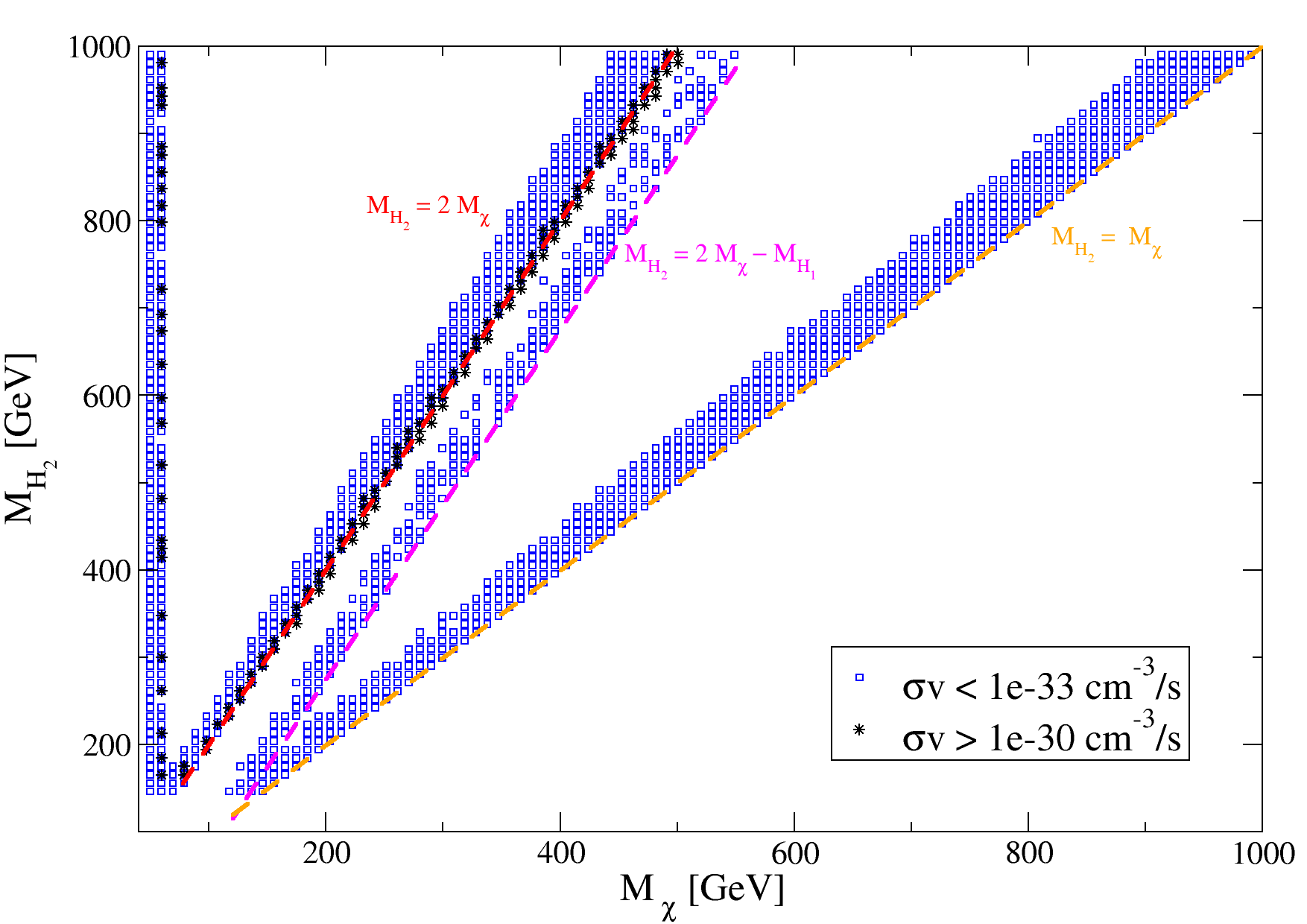}
\caption{\small Models with an unusually large (black stars) or small (blue squares) $\svt$ in the plane ($\mdm$,$\mh$). The lines $2\mdm=\mh$ (resonance), $2\mdm=M_{H_1}+\mh$ and $\mdm=\mh$ are also shown. \label{fig:fig10}}
\end{center}
\end{figure}

Figure \ref{fig:fig9} shows a scatter plot of the annihilation rate today, $\svt$, versus the dark matter mass, with different symbols according to the dominant annihilation final states. As can be seen, the annihilation rate varies over many orders of magnitude. Naively, given that a $\sv\sim 10^{-26}\mathrm{cm^3s^{-1}}$ is required in the early Universe to obtain the correct dark matter density one expects $\svt$ to naturally be today of order $ 10^{-31}\mathrm{cm^3s^{-1}}$ in this model. And many models indeed feature a $\svt$ around this value. But we also see a handful of models at low masses with a $\svt$ significantly larger than our estimate, and plenty of models with arbitrary masses featuring values of $\svt$ orders of magnitude below  it. How can we explain these deviations? The answer to this question is illustrated in figure \ref{fig:fig10}, which shows all \emph{unusual} models  in the plane ($\mdm$,$\mh$). For clarity, the lines $2\mdm=\mh$ (red), $2\mdm=\mhone+\mh$ (magenta) and $\mdm=\mh$ (orange) are also displayed. Notice that the small number of models with an annihilation cross section larger than expected (black stars) all lie at the $H_1$ or $H_2$ resonances. Models with a suppressed annihilation cross section, on the other hand, lie either at one of the two resonances or at the thresholds for $H_1H_2$ and $H_2H_2$ production. The reason $\svt$ is suppressed in this latter case is that whereas in the early Universe the dark matter particles have enough energy to cross the threshold and annihilate into these final states,  nowadays these channels are kinematically closed. This figure also emphasizes that in this model resonances and thresholds  play an important role in the calculation of the dark matter relic density. That is one of the reasons why  it is important to compute the relic density with tools such as micrOMEGAs, which automatically take into account these effects.   

Current data from indirect detection experiments such as Fermi-LAT have only recently started to exclude \emph{thermal} cross sections ($\svt \sim 3\times 10^{-26}\mathrm{cm^3s^{-1}}$) for very light dark matter particles $m\lesssim 30~\gev$ \cite{Ackermann:2012rg,Ackermann:2011wa}. They do no impose, therefore, any bounds on the models we have discussed nor are they expected to do so in the near future. The problem is that the expected gamma-ray flux from dark matter annihilation is proportional to $\svt$, which as we have seen is quite suppressed in this model. Antimatter signals \cite{Salati:2010rc}, which provide alternative ways to indirectly search for dark matter, are also proportional to $\svt$ and therefore equally suppressed. The only possibility, and a rather remote one,  to have a non-negligible indirect signal is provided by high-energy neutrinos coming from the annihilation of dark matter particles captured in the Sun \cite{Halzen:2009vu}. Let us briefly review the necessary formalism and explain why that is the case.     

\begin{figure}[tb]
\begin{center} 
\begin{tabular}{cc}
\includegraphics[scale=0.2]{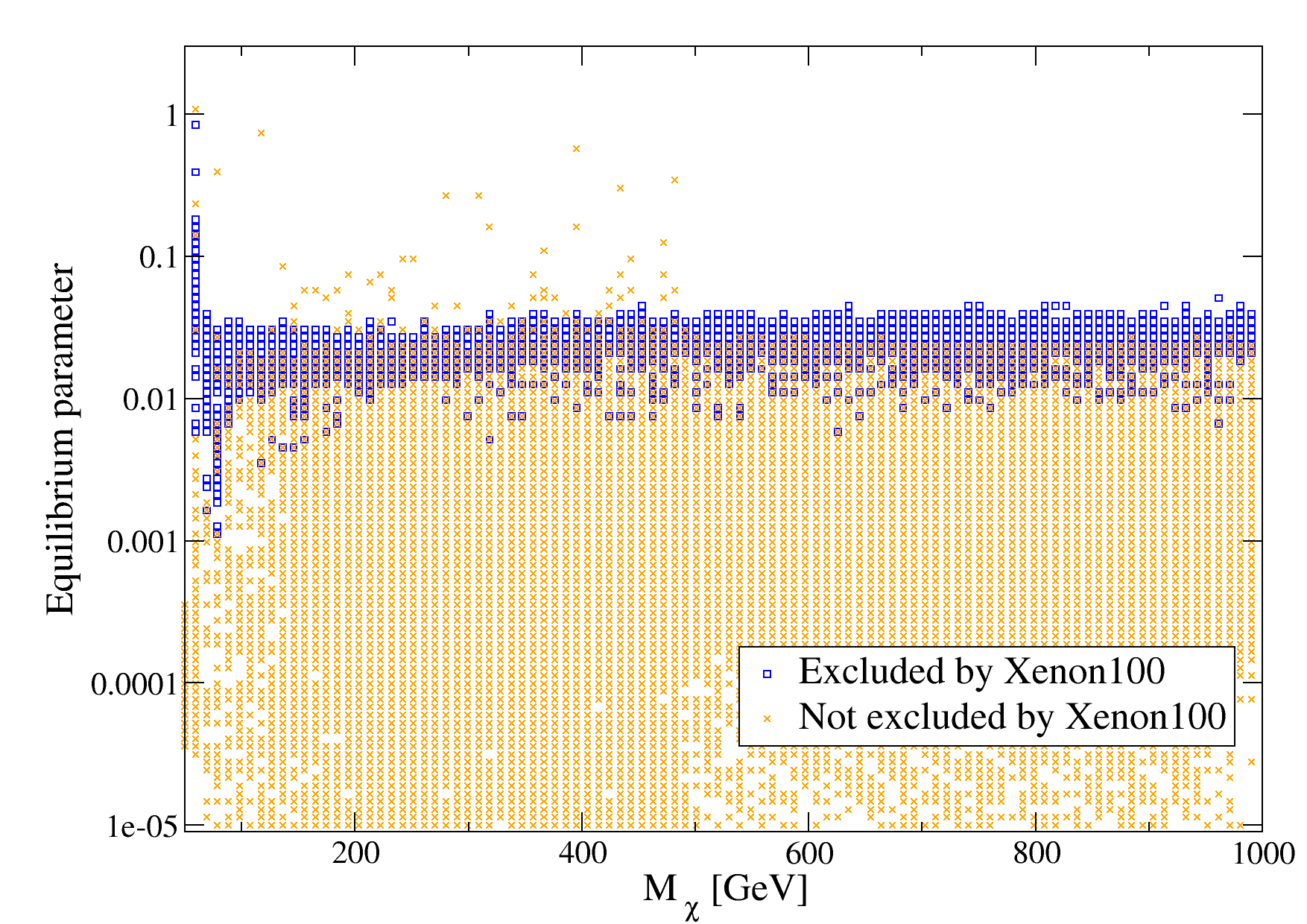} & \includegraphics[scale=0.2]{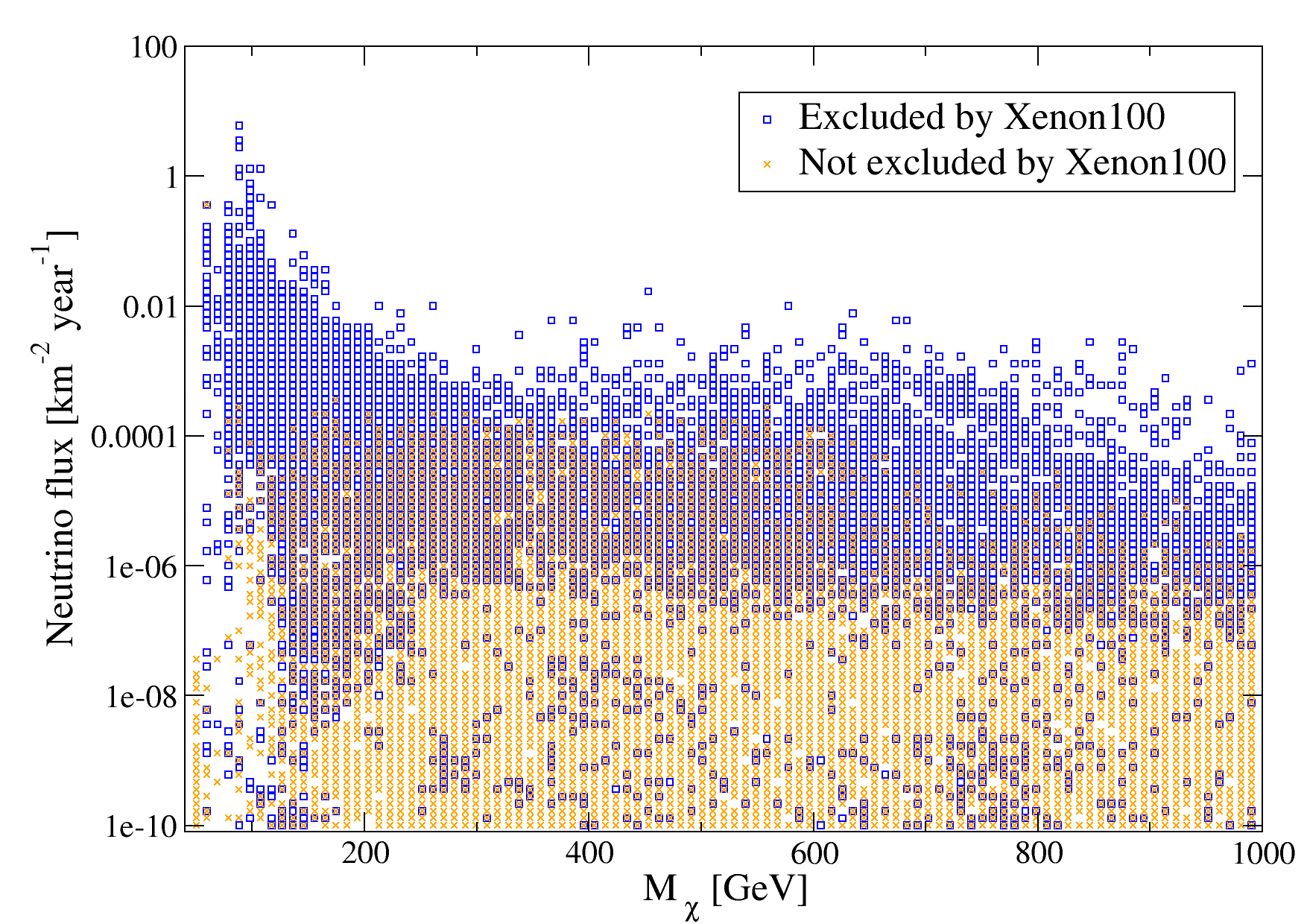}
\end{tabular}
\caption{\small A scatter plot of the  equilibrium parameter (left) and the neutrino flux (right) versus the dark matter mass.  Models which are excluded by the current XENON100 bound are shown as blue squares; those that are not are shown as orange exes.  \label{fig:fig11}}
\end{center}
\end{figure}

Dark matter particles can be captured by the Sun and then annihilate into SM particles, which in turn decay producing neutrinos that can be observed at the Earth. The equation describing the evolution of the number of dark matter particles $N_\chi$ is\footnote{Here we neglect evaporation effects, which are relevant only for very light particles.} 
\begin{equation}
 \dot{N}_\chi= C_\chi -A_{\chi\chi} N_\chi^2,
\end{equation}
where $C_\chi$ is the capture rate and $A_{\chi\chi} N_\chi^2$ is the annihilation rate of captured particles --see e.g. \cite{Belanger:2013oya}. $C_\chi$ is determined by the dark matter-nucleus scattering cross section whereas  $A_{\chi\chi}$ depends on the annihilation cross section, $\svt$. If the capture and annihilation rates are sufficiently large, that is when the equilibrium parameter --$\sqrt{C_\chi A_{\chi\chi}}t$ (with $t=4.57\times 10^9$ years for the Sun)-- is much larger than one,  equilibrium is reached and the annihilation rate of captured particles is only determined by the capture rate: $A_{\chi\chi}N_{\chi}^2=C_\chi$. This is the crucial point that gives neutrinos a small chance in spite of the suppressed $\svt$: if equilibrium is reached the number of captured dark matter particles annihilating in the Sun is determined solely by the capture rate (which in turn depends on the spin-independent direct detection cross section) and not by $\svt$. Whether equilibrium is reached or not, however, does depend on $\svt$. And  a small value of $\svt$ certainly makes it more difficult to reach equilibrium. In the singlet fermionic model, the main source of neutrinos are the decays of the $W$'s (and to a lesser extent of the $Z$'s) produced in dark matter annihilations, which as we have seen are more relevant at lower masses. The final states including $H_1$ and $H_2$ do not produce a significant signal and their contributions are not taken into account in our calculations. Figure \ref{fig:fig11} shows a scatter plot of the equilibrium parameter (left panel) versus the dark matter mass. We see that unfortunately equilibrium in the Sun is never reached in this model. In fact the equilibrium parameter hardly ever goes above  $0.1$. Consequently, the neutrino flux on earth is extremely small, as illustrated in the right panel.  It reaches at most $10$ neutrinos per $\mathrm{km^2 year}$, and that for models already excluded by XENON100. Non-excluded models generate a neutrino flux below  $10^{-4}$ neutrinos per $\mathrm{km^2 year}$.

We conclude then that, for the parity-conserving case, the indirect detection of singlet fermionic dark matter is hopeless. Being suppressed by $v^2$, the annihilation rate today is usually at least five orders of magnitude smaller than the \emph{thermal} one, and close to  resonances and thresholds it can reach even smaller values. As a result, the indirect detection via gamma rays or antimatter searches is impractical. We examined the possibility of detecting the high-energy neutrinos from dark matter particles annihilating in the Sun but found that the small value of $\svt$ prevents equilibrium from being  reached and leads to a negligible neutrino flux on Earth. Fortunately, these results are significantly modified once we move to the more general case,   $g_p\neq 0$. 

\section{The general case: $g_p\neq 0$}
\label{sec:gen}
The parity-conserving case we examined in the previous section has received the most attention in the previous literature but it does not describe the general situation within the singlet fermionic dark matter model. It may well happen, in fact, that  $g_p\neq 0$, with important implications for the calculation of the relic density and for the detection prospects of  dark matter. A non-zero value of $g_p$ yields a contribution to the dark matter annihilation rate that is not suppressed by the velocity, making it easier to satisfy the dark matter constraint and significantly improving the indirect detection prospects. The direct detection cross section, in constrast, is not modified because the $g_p$ contributions are suppressed by the square of the dark matter velocity. As a result of these two effects, the direct detection bounds  become weaker. In  \cite{LopezHonorez:2012kv}, this possibility --the so-called pseudo-scalar Higgs portal-- was mentioned as one of the three possible ways to make the model viable and consistent with the current experimental data. A quantitative analysis of this region of the parameter space, however, has not been presented before.  

In our analysis, we will use  a new scan of the entire parameter space of this model. The only novelty with respect to that of the previous section (see table \ref{tab}) is that we now allow $g_p$ to vary independently within the same range as $g_s$. First,  we will examine the parameter space that is compatible with the dark matter constraint and with direct detection bounds, and we will determine the regions that are going to be probed by the next generation of direct detection experiments. Then, we will study indirect detection and compute the expected neutrino and muon flux within this model. 
\subsection{The viable regions and  direct detection bounds}

\begin{figure}[t]
\begin{center} 
\includegraphics[scale=0.4]{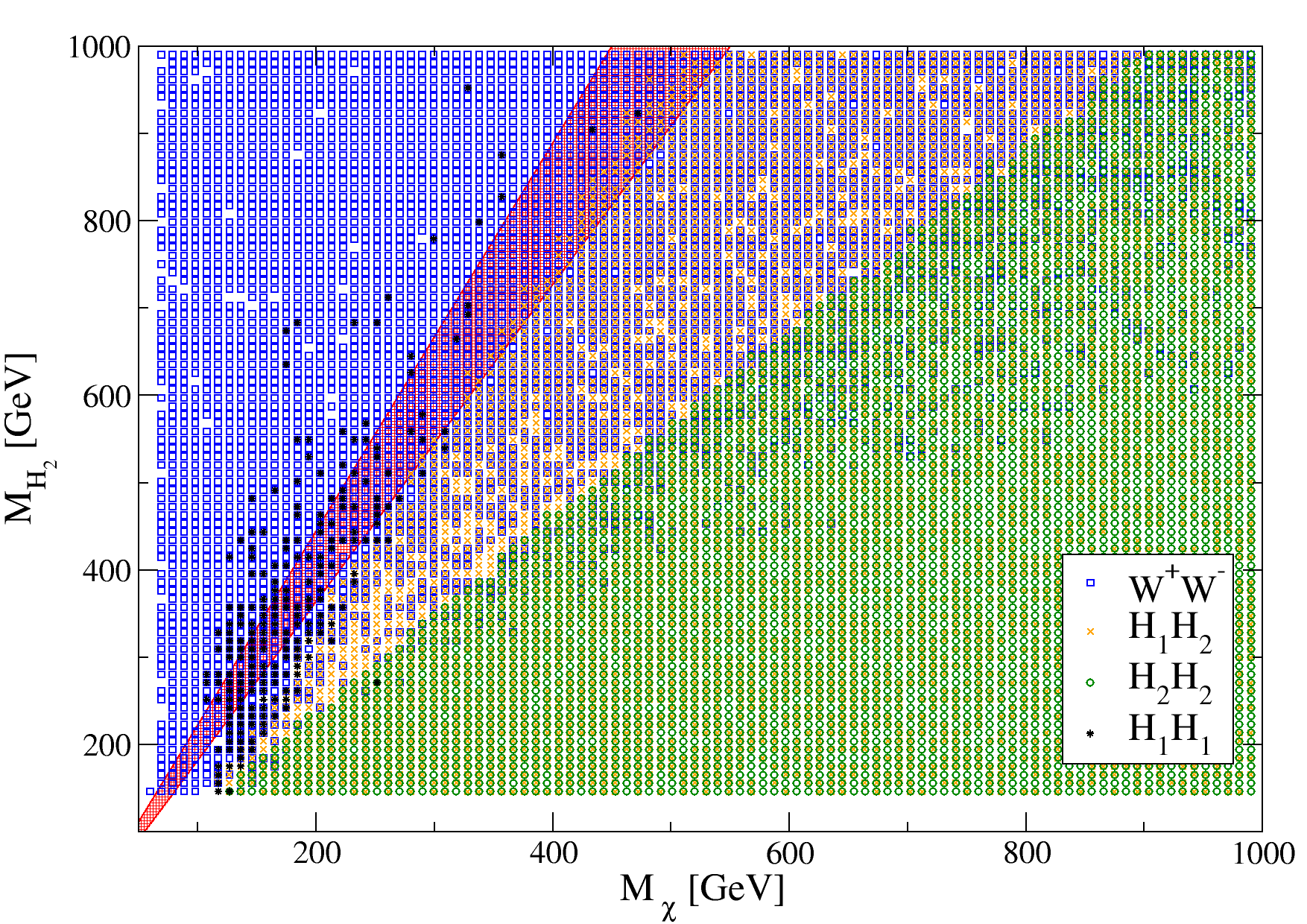}
\caption{\small Models  in the plane ($\mdm$,$\mh$) that are consistent with the dark matter constraint and with  the XENON100 bound for the general case ($g_p\neq 0$).  Different symbols are used to distinguish the dominant annihilation final states. The red band illustrates the resonance region. \label{fig:gpfig1}}
\end{center}
\end{figure}

When  $g_p\neq 0$ the dark matter annihilation rate in the early Universe, $\sv$, receives a contribution (proportional to $g_p^2$) that is independent of the velocity, making it easier to satisfy the relic density constraint. The direct detection cross section, on the other hand,  is not modified and it is still independent of $g_p$ --see equation (\ref{eq:dd}). Consequently, the correlation between these two observables is lost and the direct detection bounds become  weaker. Figure \ref{fig:gpfig1} shows, in the plane ($\mdm$,$\mh$), the models that are consistent with the dark matter constraint and with the XENON100 bound. This figure is therefore analogous to figure \ref{fig:fig8} for the parity-conserving case. As before, the models are distinguished according to the dominant annihilation final states. A crucial difference is that now we find plenty of viable models that annihilate mainly into $W^+W^-$ and with large masses scattered through the region above the resonance. Notice also that models that annihilate mainly into $H_1H_2$ are now more common in the heavy mass region, where they largely overlap with those having $H_2H_2$ as their dominant annihilation final state. Regarding the direct detection bounds, we see that, in contrast to figure \ref{fig:fig8}, there are no regions in this plane that are completely ruled out. For every value of ($\mdm$,$\mh$) it is possible to obtain the right relic density and a scattering cross section below the current limit. Future experiments will not change this result, as one can find models in the whole plane  with a $\sip$ below the expected sensitivity of XENON1T. Direct detection bounds  indeed become  weaker for $g_p\neq 0$.

\begin{figure}[t]
\begin{center} 
\includegraphics[scale=0.4]{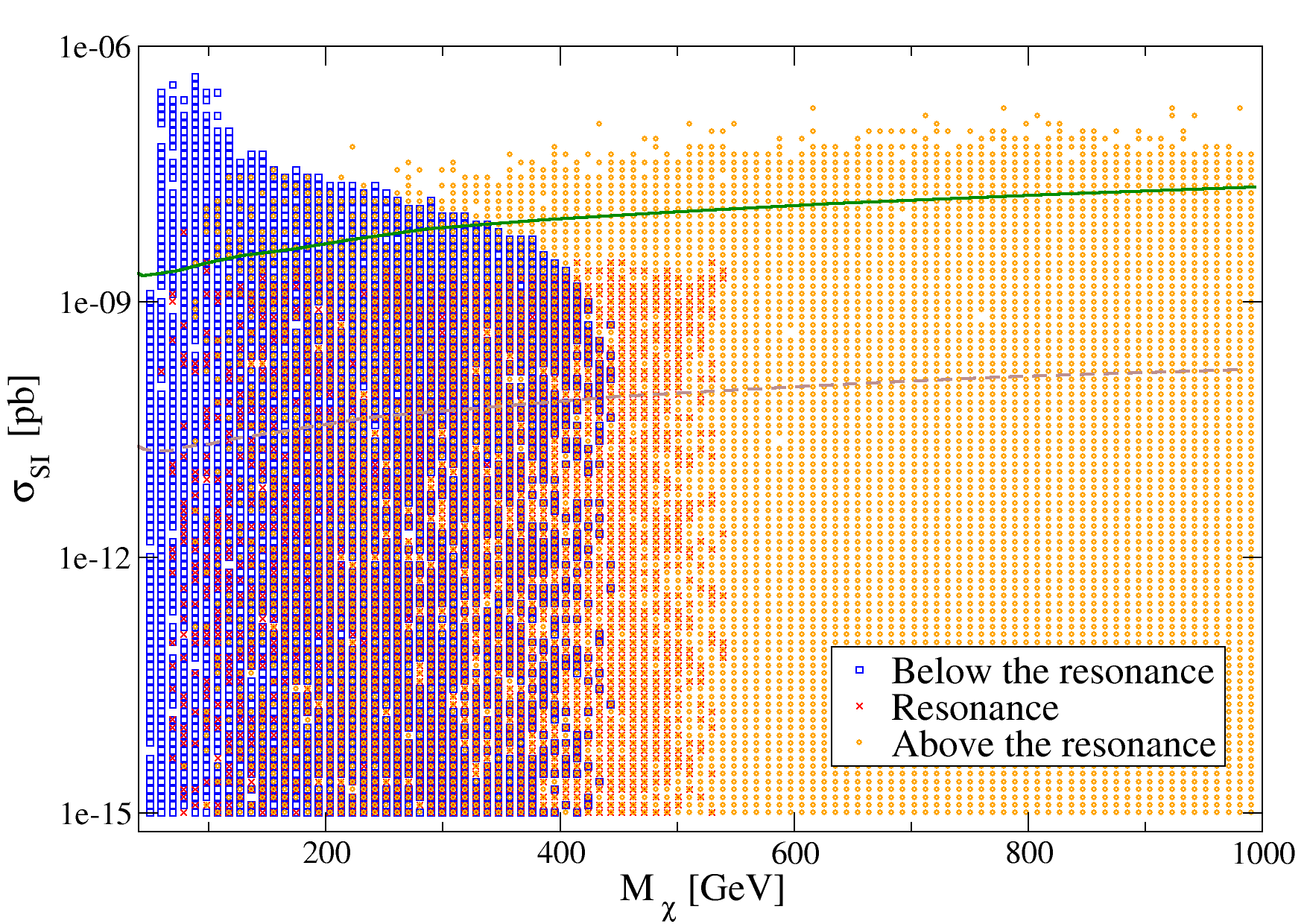}
\caption{\small A scatter plot of the spin-independent direct detection cross section versus the dark matter mass. Models above (yellow circles), below (blue sqaures) and on the resonance (red exes) are differentiated. The current bound from XENON100 is also shown (solid line) as well as the expected sensitivity of XENON1T (dashed line). \label{fig:gpfig10}}
\end{center}
\end{figure}

This conclusion is confirmed in figure \ref{fig:gpfig10}, which shows a scatter plot of the direct detection cross section versus the dark matter mass. In it, the models  are distinguished according to their position with respect to the resonance. For comparison the present bound from XENON100 (solid line) as well as the expected sensitivity of XENON1T (dashed line) are also displayed. Notice that $\sip$ can be very small for any value of $\mdm$, even below the resonance. In particular, it is no longer true that most models below the resonance will be probed by future direct detection experiments. It is clear, though, that even in this case 1-Ton experiments have the potential to exclude a significant region of the viable parameter space of this model.

\begin{figure}[t]
\begin{center} 
\includegraphics[scale=0.4]{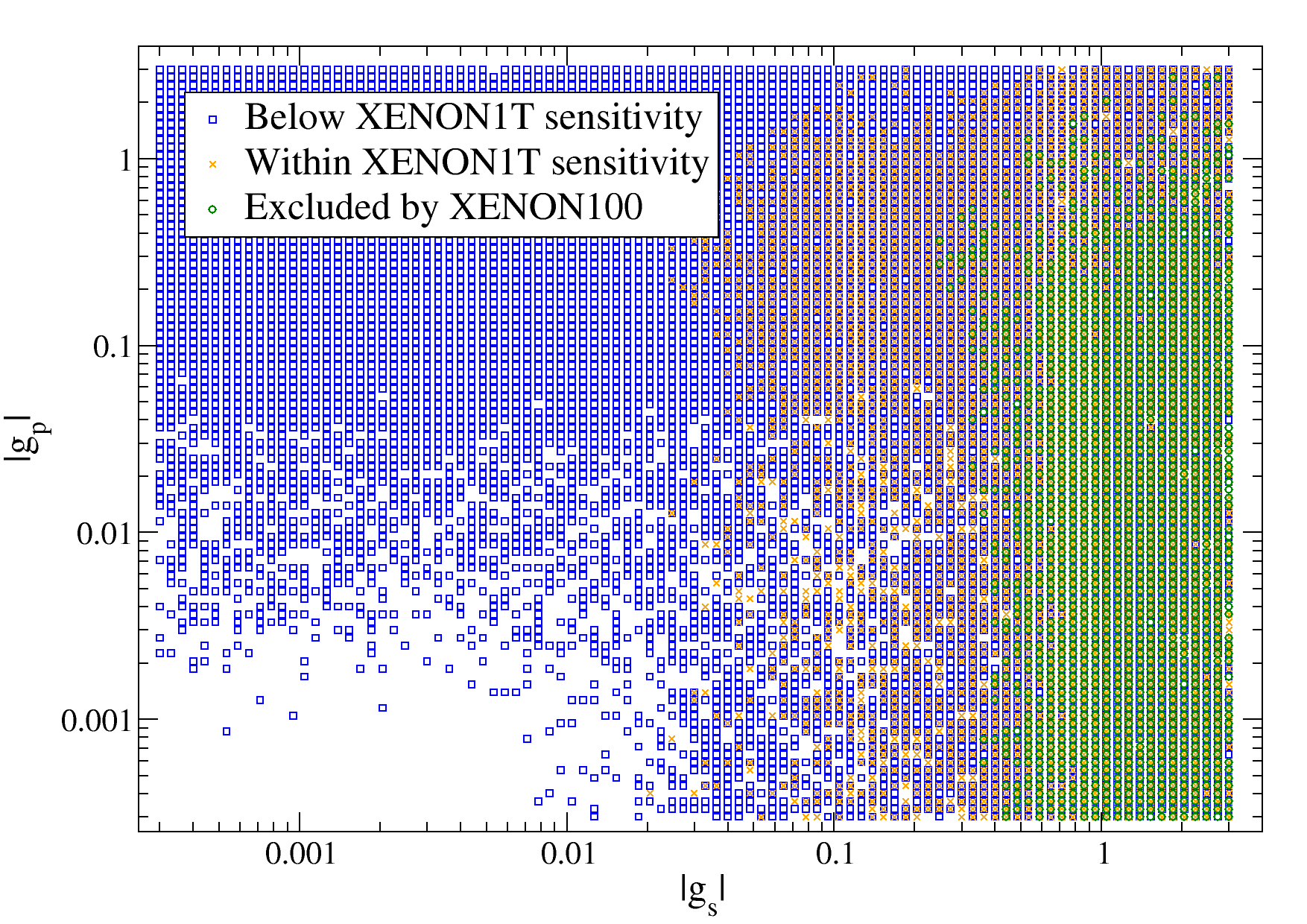}
\caption{\small Models  in the plane ($|g_s|$,$|g_p|$) that are consistent with the dark matter constraint. Different symbols are used to distinguish their direct detection status. Models that are already excluded by XENON100 are shown as green circles, those that are within the sensitivity of XENON1T with orange exes, and those with a spin-independent cross section below the sensitivity of XENON1T as blue squares. \label{fig:gpfig2}}
\end{center}
\end{figure}

Regarding the new viable regions, figure \ref{fig:gpfig2} shows the distribution of  models in the plane ($g_s$,$g_p$). First of all notice that the region where both couplings are very small is excluded as it is not possible to satisfy the relic density constraint in that case. Viable models feature $|g_p|\gtrsim 10^{-3}$ for small $g_s$, and   $|g_s|\gtrsim 10^{-2}$ for small $g_p$, their difference being due to the velocity dependence of the cross sections. In the figure, the models are distinguished according to their direct detection status, which should not depend on $g_p$. Models that are excluded by the current XENON100 bound (in green) typically feature $|g_s|\gtrsim 0.4$ whereas those  with a direct detection cross section within the XENON1T sensitivity (in orange)  are characterized by  $|g_s|\gtrsim 0.04$. Nevertheless, it is possible to find models with a $\sip$ below the current bound and below the expected  sensitivity of XENON1T over the entire viable plane. That is, neither current nor future bounds will allow, in this case, to exclude a specific range of $g_s$, let alone $g_p$.

\begin{figure}[t]
\begin{center} 
\includegraphics[scale=0.4]{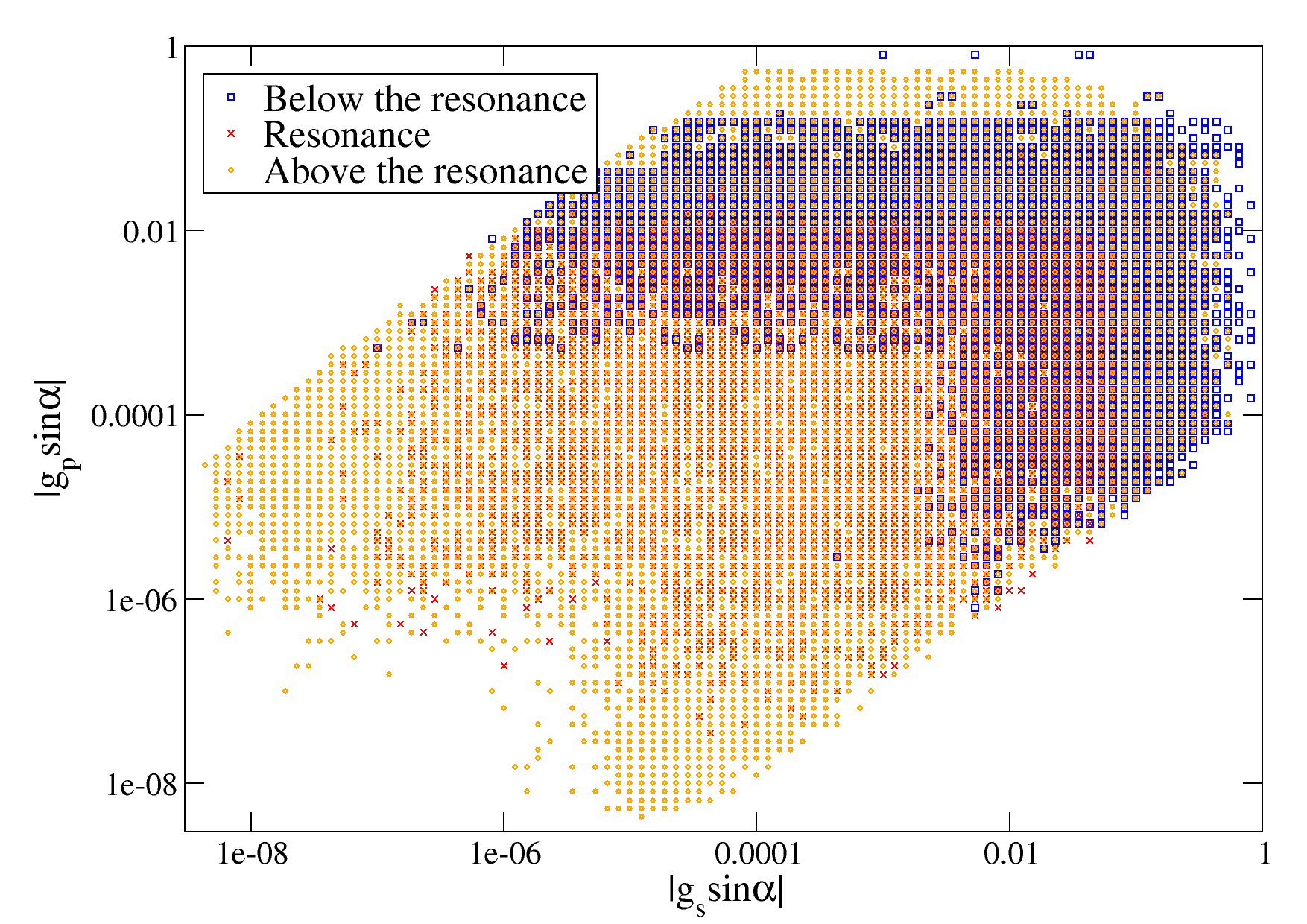}
\caption{\small Models  in the plane ($|g_s \sa|$,$|g_p\sa|$) that are consistent with the dark matter constraint and with the XENON100 bound. Models above, below and on the resonance are differentiated. \label{fig:gpfig3}}
\end{center}
\end{figure}

In the parity-conserving case, we found the phenomenology of the model to differ significantly as the resonance is crossed. Here we show that an analogous situation occurs in the case $g_p\neq 0$. Figure \ref{fig:gpfig3} shows in the plane ($|g_s \sa|$,$|g_p\sa|$) the models that are consistent with the relic density constraint and with the XENON100 bounds, differentiated according to their position with respect to the  resonance. We see that models below the resonance (blue squares) are concentrated along a narrow band, indicating that, due to the dark matter constraint, $g_s \sa$ and $g_p\sa$ cannot be simultaneously small. Indeed, $|g_s \sa|\gtrsim 0.01$ when  $|g_p \sa|$ is small whereas $|g_p \sa|\gtrsim 0.001$ for small $|g_s \sa|$. Above the resonance, on the other hand, the processes that determine the relic density become independent of $\sa$ with the result that $g_s \sa$ and $g_p\sa$ are allowed to vary over a much wider range. At the resonance both parameters are constrained to be relatively small.

We have in this way characterized the regions that are consistent with the dark matter constraint and with current direct detection bounds in the case $g_p\neq 0$. And we have examined the potential of future direct detection experiments to  further probe the parameter space of the singlet fermionic model. The introduction of a non-zero $g_p$  makes it easier to simultaneously satisfy the dark matter constraint and current direct detection bounds, relaxing the tension between these two observables.  
\subsection{Indirect Detection}
The indirect detection prospects of singlet fermionic dark matter are significantly modified once the generic case $g_p\neq 0$ is considered. The reason being that $\svt$ receives a new contribution (proportional to $g_p^2$) that is independent of the velocity, which is expected to be dominant over most of the parameter space. Here we analyze  the indirect detection signals of this model with a particular emphasis on the high-energy neutrinos from the Sun.    

\begin{figure}[tb]
\begin{center} 
\begin{tabular}{cc}
\includegraphics[scale=0.2]{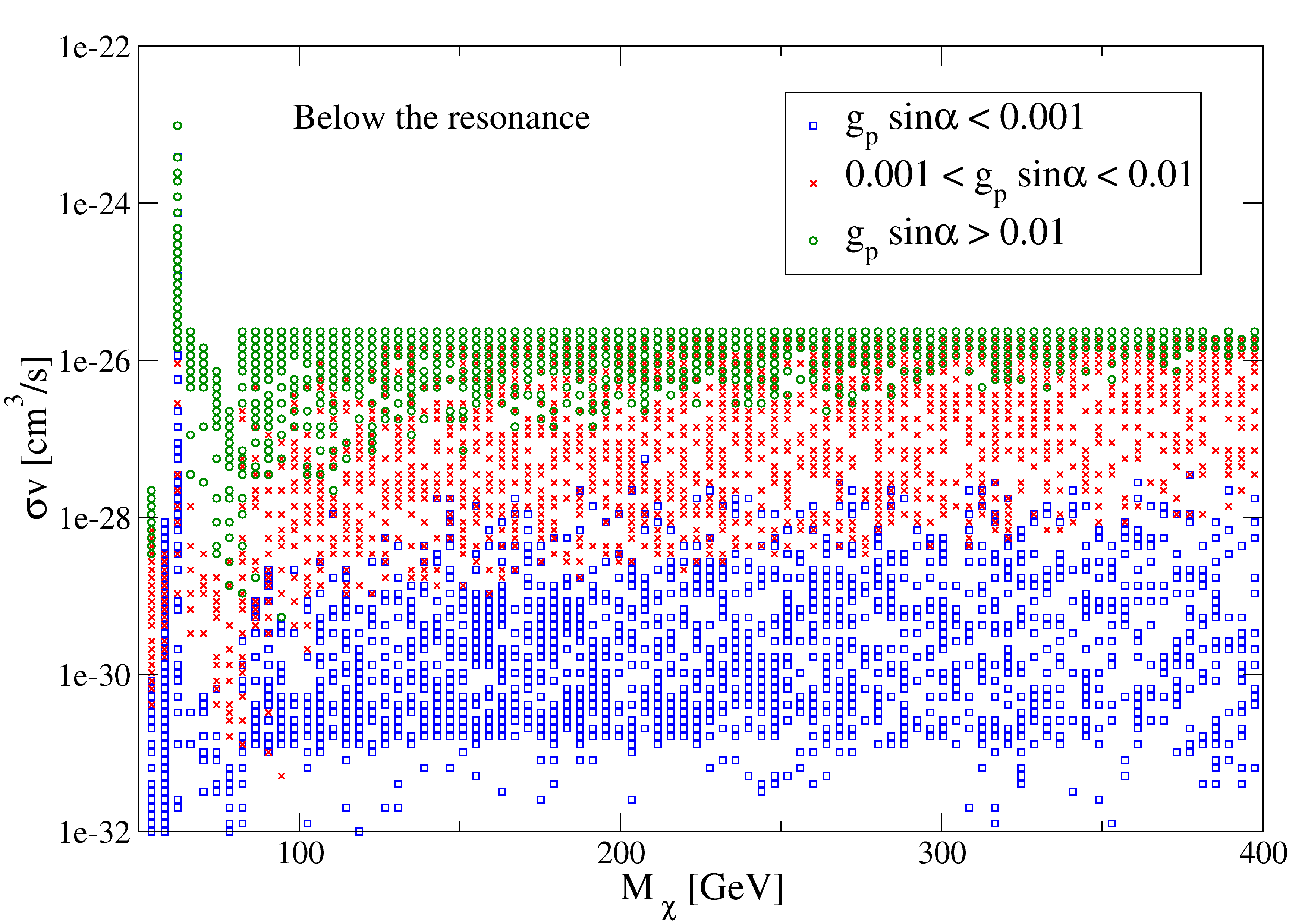} & \includegraphics[scale=0.2]{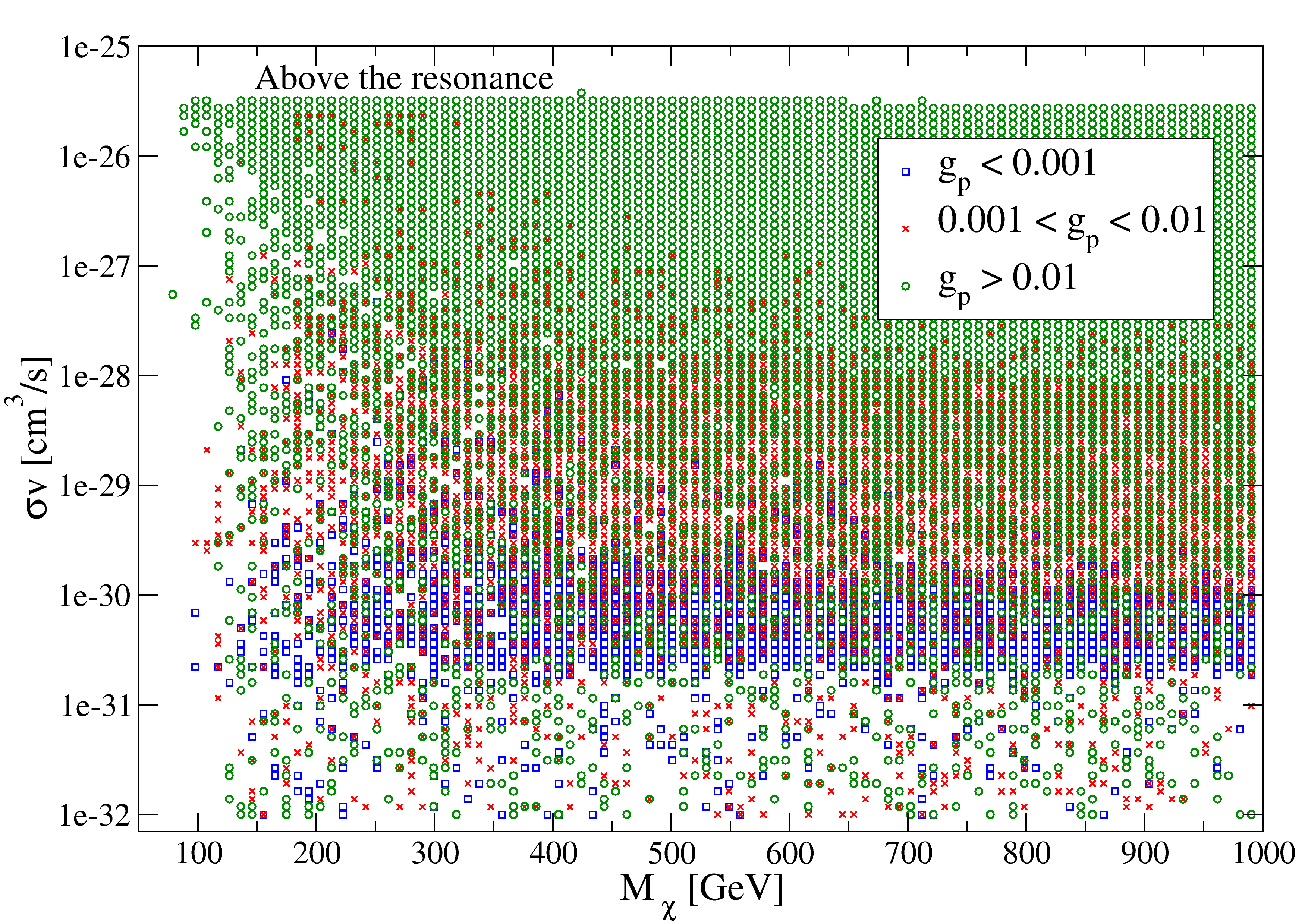}
\end{tabular}
\caption{\small A scatter plot of the  dark matter annihilation rate versus the dark matter mass for models below the resonance (left) and above it (right).  Different symbols are used to distinguished the value of $g_p\sin\alpha$ (left) or $g_p$ (right).   \label{fig:gpfig5}}
\end{center}
\end{figure}

Figure \ref{fig:gpfig5} shows a scatter plot of $\svt$ versus $\mdm$ for models below the resonance (left panel) and above it (right panel). Because $g_p\sin\alpha$ is  the parameter that determines $\svt$ for models below the resonance, different ranges for it are illustrated in the figure. As expected, $\svt$ increases with $g_p\sin\alpha$. We see that in this case $\svt$ can reach the so-called \emph{thermal} value, $\sim 3\times 10^{-26}\mathrm{cm^3s^{-1}}$, when $g_p\sin\alpha$ is sufficiently large. For models above the resonance, the crucial parameter is $g_p$ rather than $g_p\sin\alpha$. So, in the right panel, different ranges for $g_p$ are illustrated. A good correlation is observed between $g_p$ and $\svt$, and again we see that the thermal value can be reached.   

\begin{figure}[tb]
\begin{center} 
\begin{tabular}{cc}
\includegraphics[scale=0.2]{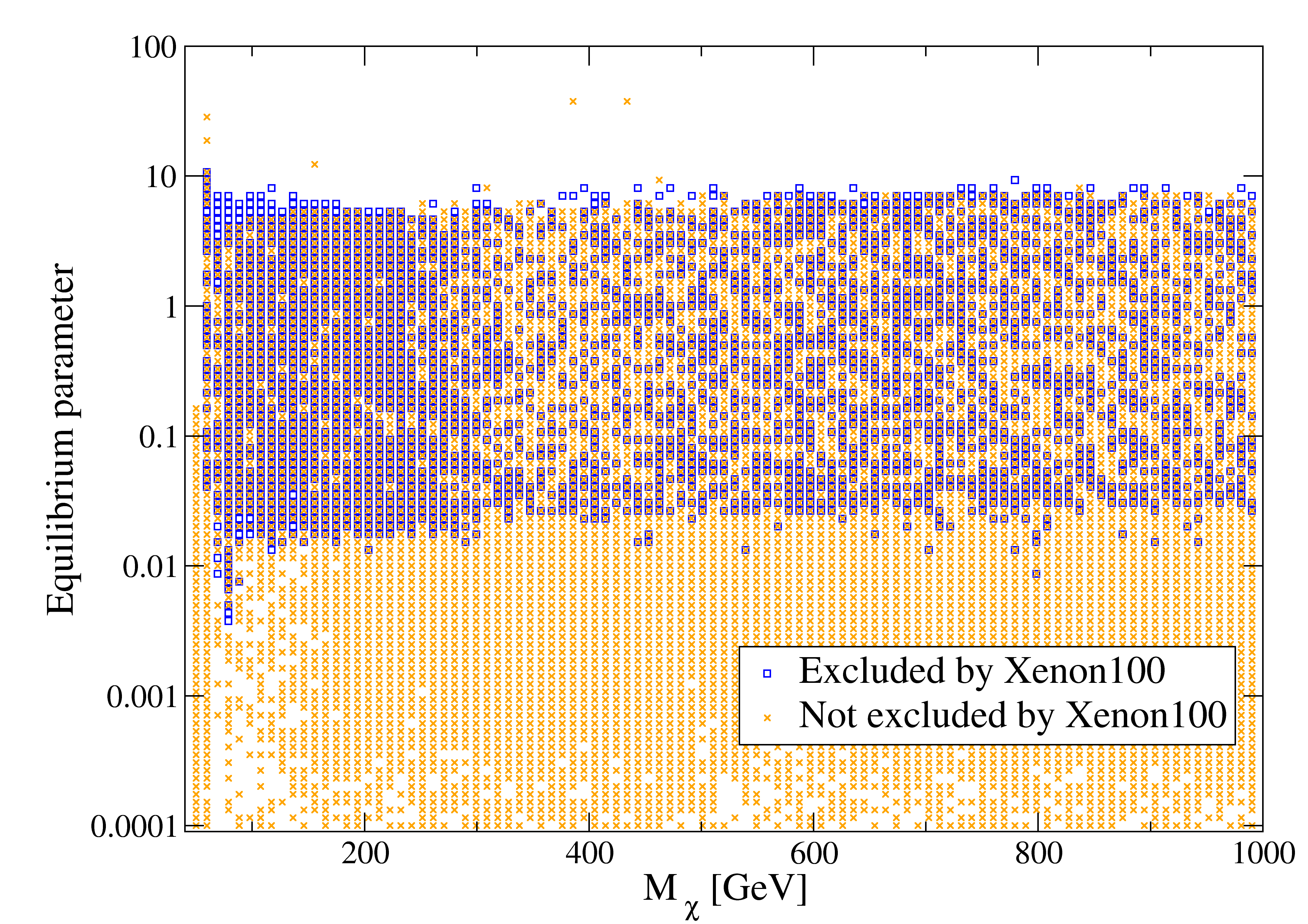} & \includegraphics[scale=0.2]{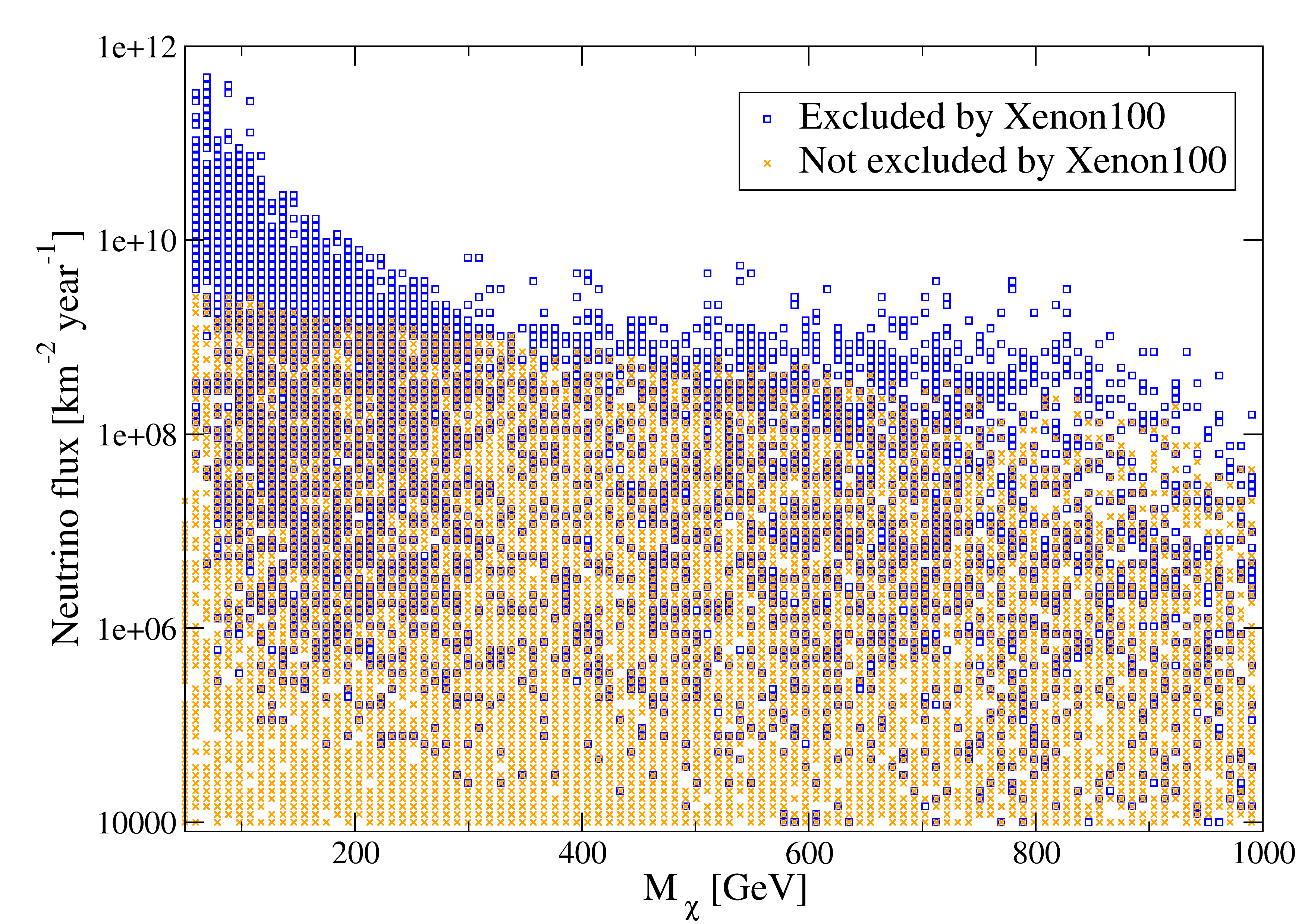}
\end{tabular}
\caption{\small Left: A scatter plot of the  equilibrium parameter versus the dark matter mass.  Right: A scatter plot of the neutrino flux versus the dark matter mass.  \label{fig:gpfig7}}
\end{center}
\end{figure}

The much larger value of $\svt$ associated with $g_p\neq 0$ implies more promising prospects for the indirect detection of singlet fermionic dark matter. Even in this case, however, we do not expect the gamma ray and antimatter channels to  provide any constraints on the parameter space of this model. As already mentioned, the Fermi-LAT data, for example, has only started to probe a thermal $\svt$ for very low masses \cite{Ackermann:2012rg,Ackermann:2011wa}. For that reason, we will focus on the indirect detection of high-energy neutrinos coming from the Sun. In this general case, the larger value of $\svt$ is expected to make equilibrium in the Sun more easily achievable than in the parity-conserving case, and, in addition, models where the dark matter particles  annihilate into $W^+W^-$ (which is the main source of neutrinos in this model) are more common and not limited to the low mass region --see figure \ref{fig:gpfig1}. As a result of these two effects,  a much larger flux of neutrinos is expected. The left panel of figure \ref{fig:gpfig7}  shows the dark matter mass versus the  equilibrium parameter. We see that this parameter can indeed be larger than one, indicating that equilibrium has been reached between the dark matter capture and annihilation processes in the Sun.   Importantly, we also learn from the figure that even though models excluded by XENON100 tend to have a larger equilibrium parameter, not all models featuring  a large one are currently excluded. That is, there is a large fraction of models, with masses in the entire range we consider, for which equilibrium is reached that are compatible with current direct detection searches. The right panel of figure \ref{fig:gpfig7} shows the resulting neutrino flux as a function of the dark matter mass.  We see that models with light dark matter candidates ($\mdm\lesssim 300~\gev$) yield the largest fluxes, which can reach values close to $10^{12}\mathrm{km^{-2}year^{-1}}$. From the figure we also learn, though,  that all such models are already  excluded by  the XENON100 data. Models not yet excluded yield a maximum neutrino flux three orders of magnitude smaller.

\begin{figure}[t]
\begin{center} 
\includegraphics[scale=0.4]{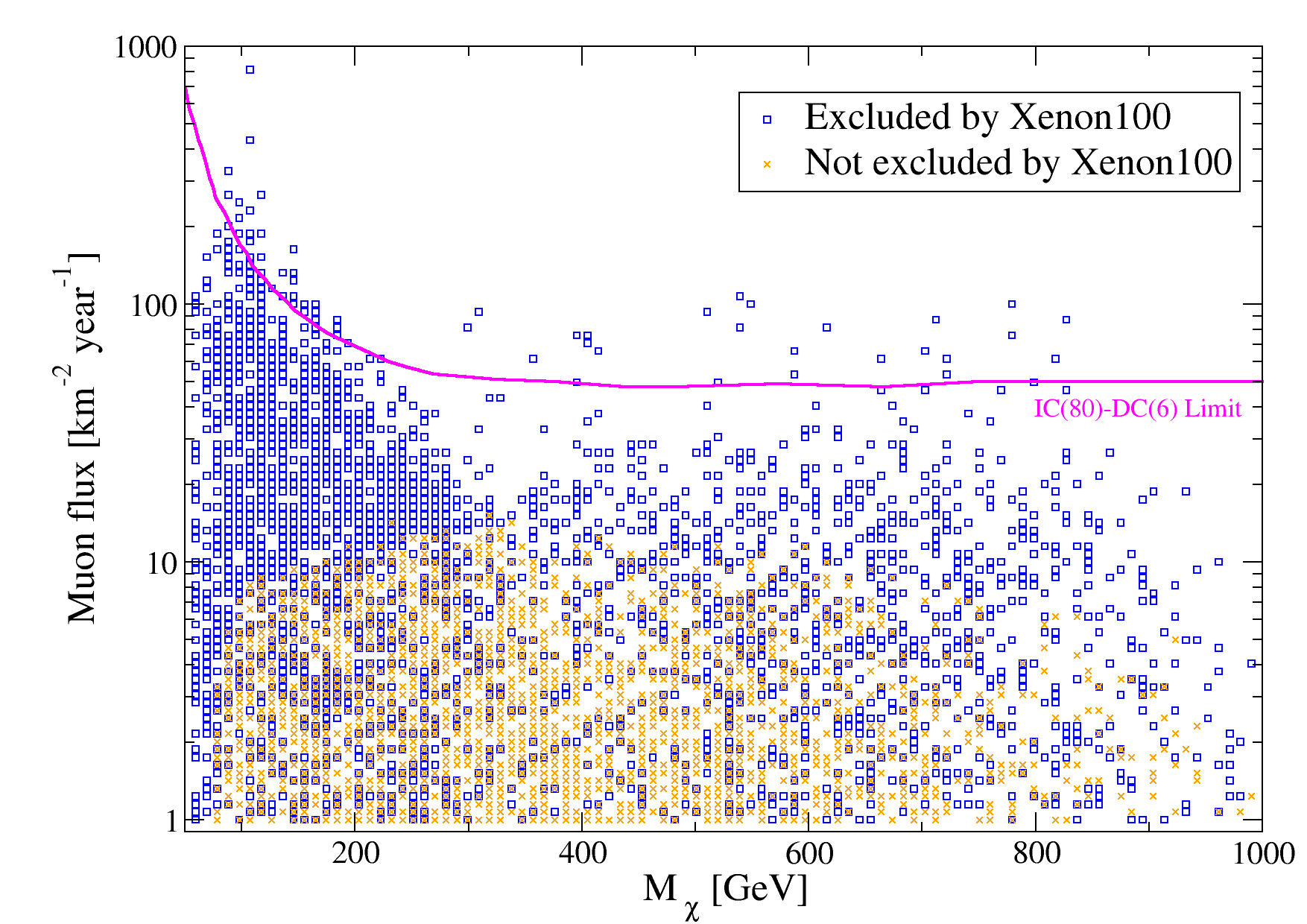}
\caption{\small The muon flux versus the dark matter mass.  Blue squares (orange exes) denote points (not) excluded by the XENON100 bound. The current bound from IceCube-DeepCore \cite{Danninger} is also shown as a solid line. \label{fig:gpfig9}}
\end{center}
\end{figure}

Neutrino telescopes such as IceCUBE \cite{IceCube:2011ab,DeYoung:2011ke,Collaboration:2011ym} detect, via Cherenkov radiation, the muons produced by neutrinos interacting with the antarctic ice.  They can set limits, therefore, on the resulting muon flux. Figure \ref{fig:gpfig9} compares the predicted muon flux for our set of models with the current bounds from IceCube-DeepCore\cite{Aartsen:2012kia, Danninger}. We see that  a handful of models are indeed ruled out by the IceCube bounds and, interestingly, they are not restricted to the low mass region, with some of them reaching $\mdm\sim 800~\gev$. Unfortunately, as illustrated in the figure, all such models are also excluded by XENON100 so we do not obtain new constraints from neutrino telescopes. One can see, in effect, that the models not currently excluded by XENON100 yield a muon flux that is at least a factor of $3$ below the current IceCube-DeepCore bound.  

\section{Conclusions}
\label{sec:con}
We have reanalyzed the singlet fermionic model of dark matter, in which the SM is extended with two additional singlet fields,  a fermion (the dark matter particle) and a scalar (which mixes with the SM Higgs). We separately considered  the parity conserving case ($g_p=0$) in section \ref{sec:par}, and the general case including parity-violating interactions ($g_p\neq0$) in section \ref{sec:gen}. The study was based on a sample of about $10^5$  random models satisfying the usual theoretical and experimental constraints, including the dark matter bound. We determined the dominant annihilation processes that set the value of the relic density, and characterized the viable  regions by projecting  the models onto different planes.  Current bounds from direct detection experiments were shown to exclude an important region of the parameter space in the parity-conserving case. Among the surviving regions, we found a new kind of indirect Higgs portal where the dark matter particles annihilate dominantly into  $H_1H_2$ rather than into $H_2H_2$. This region, which we studied in some detail, opens up  at lower dark matter masses and provides an alternative way of avoiding the direct detection bounds. The reach of future direct detection experiments was also investigated. Interestingly, they will start probing the resonance region and will be able to rule out  most of the region below the resonance. Indirect detection prospects, on the other hand, were found to be discouraging within the parity-conserving case because $\svt$  is suppressed not only by the velocity but also by resonance and thresholds effects. The neutrino signal from dark matter annihilations in the Sun turned out to be negligible due to the fact that equilibrium between capture and annihilation is never reached.   For the general case we did a similar analysis. The viable regions were determined and the impact of current and future direct detection experiments was illustrated. We showed that even if current direct detection bounds are not as constraining in this case,  future experiments will probe an important region of the parameter space. Indirect detection prospects were shown to be rather promising, with many models featuring a $\svt$ close to the thermal value. Equilibrium in the Sun can be achieved and the resulting neutrino and muon fluxes are much larger. In fact, a handful of models yields a muon flux above the current IceCube-DeepCore limit. All of them, however, are also excluded by the XENON100 bound.

\section*{Acknowledgments}
This work  is partially supported by the ``Helmholtz Alliance for Astroparticle Phyics HAP''
 funded by the Initiative and Networking Fund of the Helmholtz Association. 
\bibliographystyle{hunsrt}
\bibliography{darkmatter}

\begin{thebibliography}{10}

\bibitem{Patt:2006fw}
Brian Patt and Frank Wilczek.
\newblock {Higgs-field portal into hidden sectors}.
\newblock 2006, hep-ph/0605188.

\bibitem{Kim:2006af}
Yeong~Gyun Kim and Kang~Young Lee.
\newblock {The Minimal model of fermionic dark matter}.
\newblock {\em Phys.Rev.}, D75:115012, 2007, hep-ph/0611069.

\bibitem{Kanemura:2010sh}
Shinya Kanemura, Shigeki Matsumoto, Takehiro Nabeshima, and Nobuchika Okada.
\newblock {Can WIMP Dark Matter overcome the Nightmare Scenario?}
\newblock {\em Phys.Rev.}, D82:055026, 2010, 1005.5651.

\bibitem{Djouadi:2011aa}
Abdelhak Djouadi, Oleg Lebedev, Yann Mambrini, and Jeremie Quevillon.
\newblock {Implications of LHC searches for Higgs--portal dark matter}.
\newblock {\em Phys.Lett.}, B709:65--69, 2012, 1112.3299.

\bibitem{Pospelov:2011yp}
Maxim Pospelov and Adam Ritz.
\newblock {Higgs decays to dark matter: beyond the minimal model}.
\newblock {\em Phys.Rev.}, D84:113001, 2011, 1109.4872.

\bibitem{Kim:2008pp}
Yeong~Gyun Kim, Kang~Young Lee, and Seodong Shin.
\newblock {Singlet fermionic dark matter}.
\newblock {\em JHEP}, 0805:100, 2008, 0803.2932.

\bibitem{Baek:2011aa}
Seungwon Baek, P.~Ko, and Wan-Il Park.
\newblock {Search for the Higgs portal to a singlet fermionic dark matter at
  the LHC}.
\newblock {\em JHEP}, 1202:047, 2012, 1112.1847.

\bibitem{LopezHonorez:2012kv}
Laura Lopez-Honorez, Thomas Schwetz, and Jure Zupan.
\newblock {Higgs portal, fermionic dark matter, and a Standard Model like Higgs
  at 125 GeV}.
\newblock {\em Phys.Lett.}, B716:179--185, 2012, 1203.2064.

\bibitem{Baek:2012uj}
Seungwon Baek, P.~Ko, Wan-Il Park, and Eibun Senaha.
\newblock {Vacuum structure and stability of a singlet fermion dark matter
  model with a singlet scalar messenger}.
\newblock {\em JHEP}, 1211:116, 2012, 1209.4163.

\bibitem{Fairbairn:2013uta}
Malcolm Fairbairn and Robert Hogan.
\newblock {Singlet Fermionic Dark Matter and the Electroweak Phase Transition}.
\newblock 2013, 1305.3452.

\bibitem{Aprile:2012nq}
E.~Aprile et~al.
\newblock {Dark Matter Results from 225 Live Days of XENON100 Data}.
\newblock {\em Phys.Rev.Lett.}, 109:181301, 2012, 1207.5988.

\bibitem{Aad:2012tfa}
Georges Aad et~al.
\newblock {Observation of a new particle in the search for the Standard Model
  Higgs boson with the ATLAS detector at the LHC}.
\newblock {\em Phys.Lett.}, B716:1--29, 2012, 1207.7214.

\bibitem{Chatrchyan:2012ufa}
Serguei Chatrchyan et~al.
\newblock {Observation of a new boson at a mass of 125 GeV with the CMS
  experiment at the LHC}.
\newblock {\em Phys.Lett.}, B716:30--61, 2012, 1207.7235.

\bibitem{IceCube:2011ab}
R.~Abbasi et~al.
\newblock {IceCube - Astrophysics and Astroparticle Physics at the South Pole}.
\newblock 2011, 1111.5188.

\bibitem{Barger:2007im}
Vernon Barger, Paul Langacker, Mathew McCaskey, Michael~J. Ramsey-Musolf, and
  Gabe Shaughnessy.
\newblock {LHC Phenomenology of an Extended Standard Model with a Real Scalar
  Singlet}.
\newblock {\em Phys.Rev.}, D77:035005, 2008, 0706.4311.

\bibitem{Espinosa:2012ir}
J.R. Espinosa, C.~Grojean, M.~Muhlleitner, and M.~Trott.
\newblock {Fingerprinting Higgs Suspects at the LHC}.
\newblock {\em JHEP}, 1205:097, 2012, 1202.3697.

\bibitem{CMS:yva}
{Combination of standard model Higgs boson searches and measurements of the
  properties of the new boson with a mass near 125 GeV}.
\newblock 2013, CMS-PAS-HIG-13-005.

\bibitem{ATLAS:2013sla}
{Combined coupling measurements of the Higgs-like boson with the ATLAS detector
  using up to 25 fb$^{-1}$ of proton-proton collision data}.
\newblock 2013, ATLAS-CONF-2013-034, ATLAS-COM-CONF-2013-035.

\bibitem{Ellis:2013lra}
John Ellis and Tevong You.
\newblock {Updated Global Analysis of Higgs Couplings}.
\newblock {\em JHEP}, 1306:103, 2013, 1303.3879.

\bibitem{Chatrchyan:2013yoa}
Serguei Chatrchyan et~al.
\newblock {Search for a standard-model-like Higgs boson with a mass in the
  range 145 to 1000 GeV at the LHC}.
\newblock {\em Eur.Phys.J.}, C73:2469, 2013, 1304.0213.

\bibitem{Ade:2013lta}
P.A.R. Ade et~al.
\newblock {Planck 2013 results. XVI. Cosmological parameters}.
\newblock 2013, 1303.5076.

\bibitem{Belanger:2013oya}
G.~Belanger, F.~Boudjema, A.~Pukhov, and A.~Semenov.
\newblock {micrOMEGAs3.1: a program for calculating dark matter observables}.
\newblock 2013, 1305.0237.

\bibitem{Semenov:2010qt}
A.~Semenov.
\newblock {LanHEP - a package for automatic generation of Feynman rules from
  the Lagrangian. Updated version 3.1}.
\newblock 2010, 1005.1909.

\bibitem{Aprile:2012zx}
Elena Aprile.
\newblock {The XENON1T Dark Matter Search Experiment}.
\newblock 2012, 1206.6288.

\bibitem{Ackermann:2012rg}
M.~Ackermann et~al.
\newblock {Constraints on the Galactic Halo Dark Matter from Fermi-LAT Diffuse
  Measurements}.
\newblock {\em Astrophys.J.}, 761:91, 2012, 1205.6474.

\bibitem{Ackermann:2011wa}
M.~Ackermann et~al.
\newblock {Constraining Dark Matter Models from a Combined Analysis of Milky
  Way Satellites with the Fermi Large Area Telescope}.
\newblock {\em Phys.Rev.Lett.}, 107:241302, 2011, 1108.3546.

\bibitem{Salati:2010rc}
P.~Salati, F.~Donato, and N.~Fornengo.
\newblock {Indirect Dark Matter Detection with Cosmic Antimatter}.
\newblock 2010, 1003.4124.

\bibitem{Halzen:2009vu}
Francis Halzen and Dan Hooper.
\newblock {The Indirect Search for Dark Matter with IceCube}.
\newblock {\em New J.Phys.}, 11:105019, 2009, 0910.4513.

\bibitem{Danninger}
M.~Danninger.
\newblock {Thesis: A search for dark matter annihilations in the Sun with
  IceCube and related studies}.
\newblock \texttt{http://icecube.wisc.edu/~mda65/talks/thesis.pdf}.

\bibitem{DeYoung:2011ke}
Tyce DeYoung.
\newblock {Particle Physics in Ice with IceCube DeepCore}.
\newblock {\em Nucl.Instrum.Meth.}, A692:180--183, 2012, 1112.1053.

\bibitem{Collaboration:2011ym}
R.~Abbasi et~al.
\newblock {The Design and Performance of IceCube DeepCore}.
\newblock {\em Astropart.Phys.}, 35:615--624, 2012, 1109.6096.

\bibitem{Aartsen:2012kia}
M.G. Aartsen et~al.
\newblock {Search for dark matter annihilations in the Sun with the 79-string
  IceCube detector}.
\newblock {\em Phys.Rev.Lett.}, 110:131302, 2013, 1212.4097.

\end{thebibliography}
\end{document}